V.A. Melent'ev, PhD of tech. sci., senior researcher, A.V. Rzhanov Institute of Semiconductor Physics SB RAS, Novosibirsk (melva@isp.nsc.ru)

# FORMAL METHOD FOR THE SYNTHESIS OF OPTIMAL TOPOLOGIES OF COMPUTING SYSTEMS BASED ON THE PROJECTIVE DESCRIPTION OF GRAPHS

A deterministic method for synthesizing the interconnect topologies optimized for the required properties is proposed. The method is based on the original description of graphs by projections, on establishing the bijective correspondence of the required properties and the projection properties of the initial graph, on postulating the corresponding restrictions of modified projections and on iteratively applying these restrictions to them either until the projection system is solved and the projections of the desired graph are obtained, or until its incompatibility with the given initial conditions is revealed.

**ФОРМАЛЬНЫЙ МЕТОД СИНТЕЗА ОПТИМАЛЬНЫХ ТОПОЛОГИЙ ВЫЧИСЛИТЕЛЬНЫХ СИСТЕМ, ОСНОВАННЫЙ НА ПРОЕКТИВНОМ ОПИСАНИИ ГРАФОВ**

Предлагается детерминированный метод синтеза топологий интерконнекта, оптимизированных под требуемые свойства. Метод основан на оригинальном описании графов проекциями, на установлении биективного соответствия требуемых свойств и свойств проекций начального графа, на постулировании соответствующих ограничений модифицируемых проекций и на итеративном применении к ним этих ограничений либо до момента решения системы проекций и получения проекций искомого графа, либо до выявления ее несовместности с заданными начальными условиями.

*Ключевые слова*: топология интерконнекта; метод проективного описания графов, проекции графа, решение системы проекций и синтез оптимальных топологий.

# Введение

Эффективность параллельной реализации задачи на вычислительных системах (ВС) определяется их топологической совместимостью, оцениваемой степенью соответствия топологии задачи топологиям интерконнекта, т.е. допускаемой последними долей используемых задачей вычислительных ресурсов рассматриваемой системы [1]. Понятно при этом, что 100-процентной совместимостью для любых задач обладала бы ВС с полносвязным интерконнектом (если бы таковой мог быть реализован), и в этом случае можно было бы говорить о ее полной универсальности. Максимального использования вычислительных мощностей ВС для критически важных задач можно было бы добиться, реализуя адекватную этим задачам топологию интерконнекта. Однако "узкая направленность специализированных систем приводит к неоправданно высокой стоимости, что негативно сказывается на их эффективности - отношению стоимости к производительности" [2]. Следствием специализации суперкомпьютера на ограниченный набор задач является и существенное снижение быстродействия на задачах, топологически отличных от этого набора. Поэтому в данной работе речь идет о системах, "обладающих такими важными функциональным характеристикам сети как число ее абонентов (процессоров) и задержки передачи между ними, задаваемые диаметром сети" [3], т. е. обладающих универсальными в этом отношении топологическими свойствами: регулярностью и компактностью [4].

Общепринятой базовой детерминированной моделью топологии ВС является граф, множества вершин и ребер которого взаимно-однозначны, соответственно, множествам вычислительных модулей и линий связи между ними, а метрические характеристики (диаметр, эксцентриситет, расстояние между вершинами) пропорциональны задержкам передачи/приема единицы информации [5]. Поэтому синтез топологии ВС, оптимальной в отношении сетевых задержек, сводится к синтезу графа с метриками, определяемыми масштабностью решаемых на системе параллельных задач (вычислительными объемами и объемами обрабатываемых данных) и их информационными топологиями. В работе [6] предложена и достаточно подробно рассмотрена модель параллельных вычислений, устанавливающая формальные соответствия характеристик параллельных задач метрикам графа ВС.

В связи с этим возникает проблема синтеза графа с требуемыми для оптимальной реализации параллельных алгоритмов метрическими характеристиками. Проблема эта является *NP*-полной и решается, как правило, эмпирически либо с использованием стохастических, комбинаторных, генетических или других эвристических по сути алгоритмов, однако, во-первых, такие решения все равно достаточно трудоемки и, соответственно, времязатратны, а во-вторых, не гарантируют безусловно оптимального в отношении требуемых метрик результата. И то и другое обусловлено тем, что достоверность вероятностно-оптимального синтеза, определяется либо размером выборки, либо мощностью случайно сгенерированной первоначальной популяции, и связанная с этим недетерминированность может привести к непредсказуемым, в том числе деструктивным, последствиям в процессе функционирования системы [4]. Поэтому проблему синтеза случайных графов, впервые сформулированную в 1959 году в работах [7] и [8] мы здесь не затрагиваем.

Проблема детерминированного синтеза $s$-регулярных графов имеет достаточно давнюю историю и в большей степени относится к кубическим и клеточным графам. Так, первые полные списки кубических графов относятся к концу XIX века, когда де Врис опубликовал список всех кубических графов вплоть до 10 вершин [9], вручную исследовав конфигурации плоскостей, в которых каждая точка является пересечением двух прямых, а каждая прямая содержит три точки, при этом прямые интерпретируются как вершины графа, а точки как его рёбра [10]. Хронология последующих основополагающих публикаций в этой области такова: в 1966 г. определены все кубические графы до 12 вершин [11], в 1976 г. получены каталог всех кубических графов до 14 вершин [12], числа кубических графов до 18 вершин, списки связных регулярных графов степени 4, 5 и 6 и связных двудольных регулярных графов степени 3,4, и 5 [13], в 1986 г. сгенерированы все кубические графы на 20 вершинах [14], в 1992 г. сгенерированы кубические графы до 24 вершин [15], в 1998 г. получены кубические графы из 26 вершин [16], в 2011 г. сгенерированы все кубические графы до 32 вершин [17].

К сожалению, в этих работах ничего не говорится об асимптотической сложности разработанных для этого алгоритмов, однако приведенная здесь хронология и сравнительно небольшие для графов степени 3 значения их порядков, дают наглядное представление о сложности решаемых при этом проблем. Заметим кстати, что целью приведенных выше работ являлось получение *всех неизоморфных кубических графов* с заданным числом вершин, что изначально было обусловлено чисто практической задачей конструктивного перечисления *всех валентных* *изомеров аннуленов*, которые являются регулярными графами степени три [18].

Задача построения регулярного графа с заданным обхватом, заданной степенью и наименьшим возможным количеством вершин чрезвычайно трудна и до сих пор не имеет единого подхода и общего решения [19]. Достаточно посмотреть на известном актуализированном математическом сайте mathworld.wolfram.com страницу, посвященную клеточным графам [20], чтобы убедиться в том, что текущая таблица найденных клеточных графов далеко не полна. Это также указывает на то, что имеющихся в наличии методов и алгоритмов недостаточно, чтобы выявить наличие или невозможность построения таких графов.

Понятно, что возможность получения детерминированного решения за обозримое время, как правило, связана с выбором математического описания объекта исследования (в нашем случае — графа ВС). Приведем в этой связи гиперболический бытовой пример: для забивания гвоздя необходим инструмент. Использование в качестве инструмента ювелирного молоточка потребует нанесения по гвоздю не одной сотни, а то и тысячи ударов, — соответственным будет и затраченное на это время. Однако использование молотка, соответствующего размеру гвоздя, делает поставленную задачу легко и быстро разрешимой. Приведем более релевантный к рассматриваемой здесь тематике пример: описание логической задачи можно задать совокупностью отдельных высказываний и решить задачу последовательным их перебором. Описание той же задачи таблицей истинности или выражениями логической алгебры и использование соответствующего формального аппарата даст точное и нетрудоемкое решение за несравнимо меньшее время.

Алгоритмы анализа и синтеза графов традиционно базируются на использовании матричных или списковых описаний, задающих *бинарные отношения* вершин и ребер или дуг. Работа с такими низкоуровневыми отношениями подобна приведенным выше примерам использования неадекватных задачам инструментов, — отсюда и неполиномиальное (относительно порядка и размера графа) возрастание трудоемкости и временных затрат. В работах [21], [22] был предложен и получил развитие новый метод описания, оперирующий более крупными категориями — проекциями графа. Идея об использовании таких описаний для аналитического синтеза графов с требуемыми метриками впервые была высказана и проиллюстрирована примерами в [4] и [23]. Вследствие новизны этот метод описания графов пока не получил широкого распространения, однако его апробация в создании информационных систем фармацевтической промышленности [24], в исследованиях сетевых информационных систем [25] и в других, например, в [26-29] исследованиях подтвердила его эффективность. С библиографией исследований, основанных на этом методе описания графов, можно ознакомиться в обзоре [30].

В данной работе идея детерминированного синтеза топологий с заданными свойствами [31] нашла свое формальное воплощение в синтезе описанных в [4] компактных графов ВС, обладающих минимальным диаметром, и следовательно, наименьшей топологической задержкой информационного обмена между ветвями параллельной задачи.

## 1. Предварительные замечания и определения

В данном разделе представлены необходимые пояснения к используемой в работе оригинальной терминологии и соответствующие определения. Добавим, что речь здесь пойдет о простых (без петель и кратных ребер) неориентированных графах.

*Проекция* $P(v_j)$ графа $G(V,E)$ представляет собой многоуровневую конструкцию, на нулевом уровне которой расположена вершина $v_j \in V$ — основание, или ракурс проекции; порожденное основанием подмножество вершин первого уровня $V_{1j} \subset V$ содержит все вершины его окружения $\mathcal{N}(v_j) \setminus v_j$ без порождающей его вершины $v_j$, а $i$-й уровень ($i \geq 1$) представляет собой совокупность подмножеств вершин, каждое из которых порождено вершиной $(i-1)$-го уровня и является окружением этой вершины без вершин, непосредственно предшествующих ей в данной проекции [4]. Фрагменты проекции, порождаемые вершинами первого уровня, называем ветвями проекции.

Отношения непосредственного предшествования и порождения вершин в проекции $P(v)$ есть отношения смежности этих вершин. Число порожденных на $i$-м уровне подмножеств соответствует числу порождающего их вершин $(i-1)$-го уровня. Упорядочив вершины по их непосредственным предшествованиям и порождениям от ракурсной $v$ до $j$-й вершины $i$-го уровня $v_{ij}$, получим маршрут $M(v - v_{ij}) = (v, v_{1x}, \cdots, v_{ij})$ из вершины $v$ в вершину $v_{ij}$, при этом в невзвешенном графе номер $i$ уровня равен длине пути из ракурсной вершины $v$ в любую вершину подмножества $V_i$ находящихся на уровне $i$ вершин. Последовательность вершин, непосредственно предшествующих открывающимся скобкам от $v_{ij}$ до ракурсной вершины $v$, дает обратный маршрут $M(v_{ij} - v) = M^{-1}(v - v_{ij})$.

Если в проекции графа одна и та же вершина встречается $m$ раз, то ее экземпляр, наиболее приближенный к основанию и расположенный первым слева, считаем оригинальным, а остальные $(m-1)$ ее копий называем реплицированными, или репликами [32]. Появление реплик в проекции указывает на наличие в графе циклов. Сумма номеров уровней, находящихся в разных ветвях оригинальной вершины и ее реплики, или любых двух реплик одной и той же вершины определяет длину соответствующего цикла.

Проекция $P(v_j)$ графа $G(V,E)$ является полной, если ею определены все вершины и все ребра этого графа [4]. Вершинно полная проекция графа $G(V,E)$ содержит все вершины множества $V$, но не обязательно является полной, тогда как реберно полная проекция является полной всегда. Минимальное число уровней[1] вершинно полной проекции, порожденных ракурсной вершиной (основанием проекции), определяется эксцентриситетом основания, тогда как минимальное число уровней в полной проекции может быть на единицу большим, если хотя бы пара оригинальных вершин эксцентриситетного уровня смежна. Понятно, что все вершины и все ребра графа будут определены совокупностью даже одноуровневых его проекций, каждая из которых по определению задает все отношения смежности ее основания. С доказательствами изложенного здесь материала можно ознакомиться в работе [21].

Проиллюстрируем вышесказанное проекциями 3-мерного гиперкуба. Приведение его здесь в виде рисунка считаем избыточным в силу общеизвестности того, что смежность вершин задана их нумерацией: расстояние между вершинами гиперкуба равно двоичному расстоянию между их номерами.

$$P(0) = 0^{\left(1^{(3^{(2,7)}, 5^{(4,7)})}, 2^{(3^{(1,7)}, 6^{(4,7)})}, 4^{(5^{(1,7)}, 6^{(2,7)})}\right)} \equiv 0^{\begin{pmatrix} 1^{(3^{(2,7)}, 5^{(4,7)})}, \\ 2^{(3^{(1,7)}, 6^{(4,7)})}, \\ 4^{(5^{(1,7)}, 6^{(2,7)})} \end{pmatrix}},$$

$$P(5) = 5^{\left(1^{(0^{(2,4)}, 3^{(2,7)})}, 4^{(0^{(1,2)}, 6^{(2,7)})}, 7^{(3^{(1,2)}, 6^{(2,4)})}\right)} \equiv 5^{\begin{pmatrix} 1^{(0^{(2,4)}, 3^{(2,7)})}, \\ 4^{(0^{(1,2)}, 6^{(2,7)})}, \\ 7^{(3^{(1,2)}, 6^{(2,4)})} \end{pmatrix}}.$$

---

[1] Далее словосочетание «порожденных основанием проекции» опускаем, подразумевая, что в число уровней входят только уровни, стоящие выше основания (нулевого уровня).

Описание каждой из приведенных выше проекций дано в двух тождественных вариантах: однострочном и многострочном. Во втором варианте число строк соответствует числу ветвей проекции, или степени ракурсной вершины. Такое описание, во-первых, более компактно, а во-вторых, позволяет рассматривать отдельные ветви проекций, что будет использовано нами в дальнейшем. Заметим, что любая 3-уровневая проекция 3-куба является полной, поэтому на данном примере нетрудно убедиться в простоте и нетрудоемкости использования проекций для выявления кратчайших или заданной длины маршрутов, диаметра, эксцентриситета, обхвата и прочих циклов, а также других характеристик описанного таким образом графа.

*Компактные графы вычислительных систем.*

Одной из насущных проблем управления вычислительными системами (ВС) является критичность к времени реализации управляющих воздействий. Своевременность управляющих воздействий в процессе функционирования ВС во многом определяется компактностью топологии ее интерконнекта, обеспечивающей минимум числа транзитов межпроцессорных взаимодействий. Впервые понятие компактных графов как $s$-регулярных графов порядка $n$ с минимально возможным диаметром $d$ введено в [4], там же определено условие компактности регулярного графа, связывающее степень $s$ его вершин со значениями $n$ и $d$:

$$1 + s\sum_{i=1}^{d-1}(s-1)^{i-1} < n \leq 1 + s\sum_{i=1}^{d}(s-1)^{i-1} \qquad (1)$$

Далее удовлетворяющие этому условию графы будем называть $n(s,d)$-компактными, и задача построения оптимальных в отношении задержек топологий ВС состоит в синтезе соответствующих условию 1 графов.

## Синтез графов с требуемыми свойствами

Требуемые свойства искомого графа $G(V,E)$ совершенно однозначно отображаются в свойствах его проекций, например, число вершин $n = |V|$ графа $G$ равно числу его проекций $|P(G)|$, значение диаметра $d$ графа указывает на то, что число уровней любой его вершинно полной проекции не превышает $d$ и т. п. Таким образом, если синтезируемый граф $G$ должен обладать некоторым множеством требуемых свойств $Q(G)$, то это множество должно быть биективно и множеству свойств его проекций $Q(P(G))$: $f: Q(G) \leftrightarrow Q(P(G))$. Очевидно, что и свойства $Q(P(v_i))$ каждой проекции $P(v_i) \in P(G)$ должны быть тождественны свойствам $Q(P(G))$: $Q(P(G)) \rightarrow Q(P(v_i))$. Тогда, построив некий связный исходный граф $G'(V,E')$ порядка $n$ с меньшим чем $|E|$ числом ребер $E' < E$, свойства проекций $Q(P'(v_i))$ которого не противоречат требуемым свойствам $Q(P(G))$, и достраивая его ребрами так, чтобы это условие непротиворечивости сохранилось для каждой проекции при каждом $E'$ вплоть до $|E'| = |E|$, получим искомый граф $G(V,E)$, если таковой существует.

Покажем, например, общие свойства искомого $n$-вершинного $s$-регулярного компактного графа с заданным диаметром $d$ и соответствующие им свойства проекций.

| $Q(G)$ | $Q(P'(G))$ |
|---|---|
| Граф $G(V,E)$ — связный степени $s$ и диаметра $d$ | Каждая $d$-уровневая проекция $P(v_i) \in P(G)$ является вершинно полной и содержит все вершины из $V$ |
| Порядок графа $G(V,E)$ — $n = V$ | Число реплик $c$ в $d$-уровневой проекции — $c = 1 + s\sum_{i=1}^{d}(s-1)^{i-1} - n$. |
| Граф $G(V,E)$ не содержит кратных ребер и петель | Первый уровень каждой проекции не содержит реплик |

Если существует $n(s,d)$-компактный граф с числом вершин $n = 1 + s\sum_{i=1}^{d}(s-1)^{i-1}$, то называем его предельно $(s,d)$-компактным, и к приведенным выше свойствам $Q(P'(G))$ добавим: любая $d$-уровневая проекция такого графа не содержит реплик, и $c = 0$.

Рассмотрим простой демонстрационный пример синтеза предельно (3,2)-компактного ($s = 3$, $d = 2$) графа. Из условия (1) получим $n = |V| = 10$.

*Шаг 0.1: Построение исходной проекции.*

Построим двухуровневую проекцию $P(v_0)$ помеченного остовного дерева[1] искомого графа $G(V,E)$ соответствующую диаметру $d = 2$, пронумеровав в ней вершины $v$ графа от $v = 0$ до $|V| - 1 = 9$: ракурсной вершине (основанию) проекции этого дерева присвоим нулевое значение, остальные вершины в проекции $P(0)$ для определенности нумеруем по порядку снизу вверх и слева направо. Так как по условию данной задачи искомый граф должен быть предельно компактным, то каждая двухуровневая вершинно полная его проекция содержит только оригинальные вершины и не содержит реплик ($c = 0$).

---

[1] В зависимости от заданных свойств искомого графа $G(V,E)$ исходный граф не обязательно должен быть остовным, но он должен содержать все вершины искомого графа, т. е. быть связным суграфом.

$$P(0) = 0 \begin{pmatrix} 1^{(4,5)}, \\ 2^{(6,7)}, \\ 3^{(8,9)}. \end{pmatrix} \quad (2)$$

Подмножества вершин, заполняющих $l$-уровень $j$-ветви проекции $P(v)$ обозначим $M_{l,j}(v)$, здесь $l \in (1, ..., d), j \in (1, ..., s)$; так, подмножества первого уровня — $M_{1,1}(0) = (1)$, $M_{1,2}(0) = (2)$, $M_{1,3}(0) = (3)$ — содержат всего по одному элементу, подмножества второго — $M_{2,1}(0) = (4,5)$, $M_{2,2}(0) = (6,7)$, $M_{2,3}(0) = (8,9)$ — содержат по два элемента. Аналогично при поиске, например, предельно (3,3)-компактного графа, где $n = 22$, третий уровень проекции $P(0)$ составляли бы подмножества $M_{3,1}(0) = (10,11,12,13)$, $M_{3,2}(0) = (14,15,16,17)$, $M_{l3,3}(0) = (18,19,20,21)$.

Прежде чем перейти непосредственно к построению проекций и перечней запрещенных в них вакансий построим начальную таблицу смежностей вершин исходного графа:

**Таблица 1.** Таблица смежностей из исходной проекции $P(0)$

| | 0 | 1 | 2 | 3 | 4 | 5 | 6 | 7 | 8 | 9 |
|---|---|---|---|---|---|---|---|---|---|---|
| **0** | 0 | **1** | **2** | **3** | 4 | 5 | 6 | 7 | 8 | 9 |
| **1** | **0** | 1 | 2 | 3 | **4** | **5** | 6 | 7 | 8 | 9 |
| **2** | **0** | 1 | 2 | 3 | 4 | 5 | **6** | **7** | 8 | 9 |
| **3** | **0** | 1 | 2 | 3 | 4 | 5 | 6 | 7 | **8** | **9** |
| **4** | 0 | **1** | 2 | 3 | 4 | 5 | 6 | 7 | 8 | 9 |
| **5** | 0 | **1** | 2 | 3 | 4 | 5 | 6 | 7 | 8 | 9 |
| **6** | 0 | 1 | **2** | 3 | 4 | 5 | 6 | 7 | 8 | 9 |
| **7** | 0 | 1 | **2** | 3 | 4 | 5 | 6 | 7 | 8 | 9 |
| **8** | 0 | 1 | 2 | **3** | 4 | 5 | 6 | 7 | 8 | 9 |
| **9** | 0 | 1 | 2 | **3** | 4 | 5 | 6 | 7 | 8 | 9 |

Здесь номер строки соответствует номеру вершины, номера смежных ей вершин (уже известных вершин ее окружения) даны в полужирном начертании, например, для вершины (строки) **0** — это вершины **1**, **2**, и **3** ($\mathcal{N}(0) = \{1,2,3\}$), для вершины **3** — вершины **0**, **8**, **9** ($\mathcal{N}(3) = \{0,8,9\}$) и т. д. Окружения вершин из $\{0,1,2,3\}$ изначально полностью определены проекцией $P(0)$ — число $f(v_i)$ «жирных» (fat) вершин в соответствующих $v_i$ строках таблицы равно заданной степени $s$ искомого графа, в рассматриваемом здесь случае — трем: $f(0) = f(1) = f(2) = f(3) = 3$. Поэтому вакансии[1] $v(v_i)$ в этих строках отсутствуют ($v(0) = v(1) = v(2) = v(3) = 0$), и остальные вершины в них скрыты (имеют серое начертание), т. е. запрещены для использования в отношениях смежности вершин соответствующих строк таблицы. Вершины с 4-й по 9-ю имеют в таблице лишь по одной из необходимых $f(v_i) = 3$ вершин, их окружения $\mathcal{N}(4) = \mathcal{N}(5) = \{1,*,*\}$, $\mathcal{N}(6) = \mathcal{N}(7) = \{2,*,*\}$, $\mathcal{N}(8) = \mathcal{N}(9) = \{3,*,*\}$, — здесь символом "*" обозначены вакансии. Вершины в строке $v_i$ таблицы, оставшиеся "черными" (black) — $b(v_i)$-вершины, составляют подмножество $B(v_i)$ вершин, претендующих на $v(v_i) = s - f(v_i)$ ее вакансий смежности.

Диагональные номера вершин в таблице скрыты, так как искомый граф по условию не имеет кратных ребер и петель. Как видим, таблица диагонально симметрична, это также определяется условием задачи — неориентированностью выстраиваемого графа.

Число $C(v_i)$ возможных на шаге комбинаций заполнения $v(v_i)$ вакансий претендующими на них $b(v_i)$-вершинами этой же строки равно числу сочетаний $C_{b(v_i)}^{v(v_i)}$. Если рассматривать строки таблицы сверху вниз, то из числа $b(v_i)$ претендующих на вакансии вершин $v_i$-строки следует исключить размещенные под диагональю (помеченные подчеркиванием) $\underline{b(v_i)}$-вершины, как уже вошедшие в комбинации предшествующих строк. Тогда число сочетаний по $v(v_i)$ из $b(v_i) - \underline{b(v_i)}$ составит $C_{b(v_i)-\underline{b(v_i)}}^{v(v_i)}$, а общее число возможных комбинаций $C = \sum_{i=0}^{n} C_{b(v_i)-\underline{b(v_i)}}^{v(v_i)}$. Перед выполнением шага 0.1 в рассматриваемом примере, когда не задано еще ни одного ребра, для любой $v_i \in V$ число вакансий $v(v_i) = 3$, а число число претендующих на них вершин $b(v_i) = 9$, при этом число вариантов $C_0$ построения искомого графа составит $C_0 = \sum_{i=0}^{9} C_9^3 = 210$, и мы имеем 210 вариантов построения графа степени 3 из 10 вершин. Выполнив шаг 0.1 и построив остов искомого графа в соответствии с таблицей 1 мы сокращаем число возможных вариантов с первоначальных $C_0 = 210$ до $C_{0.1} = \sum_{i=4}^{9} C_{b(v_i)-\underline{b(v_i)}}^{2} = 10 + 6 + 3 + 1 = 20$.

---

[1] Вакансиями называем помеченные символом "*" пустые места в проекции. Число вакансий $v(v_i)$ в строке $v_i$ таблицы смежностей определяется разницей между заданной степенью $s$ и числом "жирных" вершин в этой строке.

*Шаг 0.2: Построение множества проекций каркаса.*

Используя проекцию $P(0)$, построим остальные проекции $P'(v_i)$, $0 < v_i < |V|$ (здесь штрих в $P'(v_i)$ указывает на ее незавершенность: в ней существуют вакансии, обозначенные символом "*"). Ниже каждой проекции разместим множество $V$ вершин, в котором уже упомянутые в рассматриваемой проекции вершины скроем (пометим серым цветом), их множество обозначим $\bar{M}*(v_i)$. Тогда черными в этом множестве останутся лишь не упомянутые в проекции $P'(v_i)$ вершины множества $M*(v_i) = V \setminus \bar{M}*(v_i)$, они же — соискатели вакансий. Заметим, что в случае построения предельно компактного графа число вакансий равно числу вершин-соискателей: $c(v_i) = |M*(v_i)|$.

Проекции $P'(v_i)$, за исключением «нулевой» $P(0)$, первоначально не являются остовными и становятся таковыми лишь в процессе построения искомого графа: если такой граф будет найден, то множества $M*(v_i)$ во всех проекциях $P(v_i)$ станут пустыми (штрих в обозначении проекции здесь опущен, так как в таком случае $P'(v_i) \equiv P(v_i)$). Справа от ветвей и уровней каждой проекции поместим подмножества вершин, запрещенных для замещения вакансий этих ветвей и их уровней. Если вакансии взятых для рассмотрения уровня и ветви отсутствуют, помечаем это символом «≡». Вершины, окружения которых полностью определены изначально (как в $P(0)$) или в процессе построения (см., например, $P'(1)$), вакансий смежности не имеют.

Ниже приведены $P(0)$ и полученные из нее остальные проекции $P'(v_i) | 0 < v_i \leq 9$ каркаса искомого предельного (3,2)-компактного графа, с соответствующими каждой ветви проекции и ее уровням подмножествами $\bar{M}_{l,j}*(v_i)$ вершин, запрещенных для заполнения вакансий в $l$-й ветви и на $j$-м уровне. Изначально эти подмножества равны подмножеству $\bar{M}*(v_i)$ «серых» вершин, извлекаемому из приведенного ниже каждой проекции множества вершин $V = \bar{M}*(v_i) \cup M*(v_i)$. По результатам рассмотрения каждой проекции корректируем таблицу смежности и вакансий. Например, из $P'(1)$ видим, что вершины 4 и 5 не могут быть смежны вершинам с нулевой по пятую, поэтому корректируем четвертую и пятую строки таблицы, скрывая в них, соответственно, пятую и четвертую вершины, не затрагивая при этом "жирную" вершину **1**, хоть и попавшую в этот интервал (0-5), но уже занявшую одну из трех (степень $s$-регулярного графа $s = 3$ по условию задачи) вакансий смежности. Аналогично поступая с остальными проекциями, модифицируем таблицу.

Полученная в результате модификации таблица 2 приведена сразу после проекций, причем для улучшения визуального восприятия текущих ее изменений относительно предшествующей таблицы 1, вершины, удаляемые из состава претендующих на вакансии "черных" вершин, выделены двойным их зачеркиванием с тонированием соответствующих ячеек. На последующих шагах для удобства восприятия текущих, полученных на каждом шаге изменений, поступим аналогично. На следующем шаге эти изменения уже не выделяем, и они становятся обычными — "серыми". Аналогично помечаем изменения подмножеств $M^*(v_i)$ "черных" вершин в строках, расположенных ниже каждой проекции $P'(v_i)$. Размещенные справа от проекций подмножества вершин, запрещенных на уровнях каждой ветви, измененные на каждом шаге, начиная с первого, помечаем подчеркиванием.

$$0 \begin{pmatrix} 1^{(4,5)}, \\ 2^{(6,7)}, \\ 3^{(8,9)}. \end{pmatrix} \begin{pmatrix} \equiv^= \\ \equiv^= \\ \equiv^= \end{pmatrix},$$
$0,1,2,3,4,5,6,7,8,9;$

$$1 \begin{pmatrix} 0^{(2,3)}, \\ 4^{(*,*)}, \\ 5^{(*,*)}. \end{pmatrix} \begin{pmatrix} \equiv^= \\ \equiv^{0-5} \\ \equiv^{0-5} \end{pmatrix},$$
$0,1,2,3,4,5,6,7,8,9;$

$$2 \begin{pmatrix} 0^{(1,3)}, \\ 6^{(*,*)}, \\ 7^{(*,*)}. \end{pmatrix} \begin{pmatrix} \equiv^= \\ \equiv^{0-3,6,7} \\ \equiv^{0-3,6,7} \end{pmatrix},$$
$0,1,2,3,4,5,6,7,8,9;$

$$3 \begin{pmatrix} 0^{(1,2)}, \\ 8^{(*,*)}, \\ 9^{(*,*)}. \end{pmatrix} \begin{pmatrix} \equiv^= \\ \equiv^{0-3,8,9} \\ \equiv^{0-3,8,9} \end{pmatrix},$$
$0,1,2,3,4,5,6,7,8,9;$

$$4 \begin{pmatrix} 1^{(0,5)}, \\ *^{(*,*)}, \\ *^{(*,*)}. \end{pmatrix} \begin{pmatrix} \equiv^= \\ 0-5^{0,1,4,5} \\ 0-5^{0,1,4,5} \end{pmatrix},$$
$0,1,2,3,4,5,6,7,8,9;$

$$5 \begin{pmatrix} 1^{(0,4)}, \\ *^{(*,*)}, \\ *^{(*,*)}. \end{pmatrix} \begin{pmatrix} \equiv^= \\ 0-5^{0,1,4,5} \\ 0-5^{0,1,4,5} \end{pmatrix},$$
$0,1,2,3,4,5,6,7,8,9;$

$$6 \begin{pmatrix} 2^{(0,7)}, \\ *^{(*,*)}, \\ *^{(*,*)}. \end{pmatrix} \begin{pmatrix} \equiv^= \\ 0-3,6,7^{0,2,6,7} \\ 0-3,6,7^{0,2,6,7} \end{pmatrix},$$
$0,1,2,3,4,5,6,7,8,9;$

$$7 \begin{pmatrix} 2^{(0,6)}, \\ *^{(*,*)}, \\ *^{(*,*)}. \end{pmatrix} * \begin{pmatrix} \equiv^= \\ 0-3,6,7^{0,2,6,7} \\ 0-3,6,7^{0,2,6,7} \end{pmatrix},$$
$0,1,2,3,4,5,6,7,8,9;$

$$8 \begin{pmatrix} 3^{(0,9)}, \\ *^{(*,*)}, \\ *^{(*,*)}. \end{pmatrix} \begin{pmatrix} \equiv^= \\ 0-3,8,9^{0,3,8,9} \\ 0-3,8,9^{0,3,8,9} \end{pmatrix},$$
$0,1,2,3,4,5,6,7,8,9;$

$$9 \begin{pmatrix} 3^{(0,8)}, \\ *^{(*,*)}, \\ *^{(*,*)}. \end{pmatrix} \begin{pmatrix} \equiv^= \\ 0-3,8,9^{0,3,8,9} \\ 0-3,8,9^{0,3,8,9} \end{pmatrix},$$
$0,1,2,3,4,5,6,7,8,9;$

**Таблица 2.** Таблица смежностей для множества проекций $\{P(v_i) | v_i \in V\}$

| | 0 | 1 | 2 | 3 | 4 | 5 | 6 | 7 | 8 | 9 |
|---|---|---|---|---|---|---|---|---|---|---|
| **0** | 0 | **1** | **2** | **3** | 4 | 5 | 6 | 7 | 8 | 9 |
| **1** | **0** | 1 | 2 | 3 | **4** | **5** | 6 | 7 | 8 | 9 |
| **2** | **0** | 1 | 2 | 3 | 4 | 5 | **6** | **7** | 8 | 9 |
| **3** | **0** | 1 | 2 | 3 | 4 | 5 | 6 | 7 | **8** | **9** |
| **4** | 0 | **1** | 2 | 3 | 4 | ~~5~~ | 6 | 7 | 8 | 9 |
| **5** | 0 | **1** | 2 | 3 | 4 | 5 | 6 | 7 | 8 | 9 |
| **6** | 0 | 1 | **2** | 3 | 4 | 5 | 6 | ~~7~~ | 8 | 9 |
| **7** | 0 | 1 | **2** | 3 | 4 | 5 | ~~6~~ | 7 | 8 | 9 |
| **8** | 0 | 1 | 2 | **3** | 4 | 5 | 6 | 7 | 8 | ~~9~~ |
| **9** | 0 | 1 | 2 | **3** | 4 | 5 | 6 | 7 | ~~8~~ | 9 |

Из таблицы 2 видно, что в результате выполнения шага 0.2 число претендующих на вакансии вершин сократилось с $C_{0.1} = 20$ до $C_{0.2} = \sum_{i=4}^{9} C^2_{b(v_i) - \underline{b(v_i)}} = \binom{4}{2} + \binom{4}{2} + \binom{2}{2} + \binom{2}{2} = 14$.

*Шаг 1: Выбор и добавление в исходный граф ребра 4-6.*

Следующий этап построения искомого графа состоит в выборе и заполнении одной из незаполненных в таблице вакансий смежности. Выбираем произвольную вершину (строку таблицы) с $f(v_i) < s$, и анализируем их. Строки 0-3 вакансий не имеют — $v(0) = v(1) = v(2) = v(3) = 0$, строки 4-9 имеют по одной "жирной" вершине — $f(4) = \ldots = f(9) = 1$, т. е. имеют по две вакансии с четырьмя претендующими на них "черными" вершинами — $v(4) = \ldots = v(9) = 2$, $b(4) = \ldots = b(9) = 4$. Выбираем для определенности верхнюю из этих строк — четвертую. Итак, на две вакансии в четвертой строке претендуют вершины с 6-й по 9-ю — $B(4) = \{6,7,8,9\}$, причем соответствующие вершинам из $B(4)$ строки 6, 7, 8 и 9 имеют две отличающиеся конфигурации. *Конфигурацией* строки $v_i$ для графа $G'(V,E')$ будем называть кортеж $C'(v_i) = \langle \mathbf{N'}(v_i), N'(v_i) \rangle$ из двух непересекающихся подмножеств, первое из которых — упорядоченное по номеру множество $\mathbf{N'}(v_i)$ выделенных полужирным шрифтом вершин, смежных вершине $v_i$, а второе — также упорядоченное по номеру множество $N'(v_i)$ вершин с обычным шрифтом, претендующих на свободные вакансии смежности вершине $v_i$, т. е. потенциально смежных ей. Сравнивая конфигурации строк с номерами черных вершин из $N'(4)$, т. е. строк с шестой по девятую, заметим, что шестая и седьмая строки имеют одинаковые конфигурации — $C'(6) = C'(7) = \langle(2), (4,5,8,9)\rangle$, но эти конфигурации отличаются от также одинаковых конфигураций восьмой и девятой строк: $C'(8) = C'(9) = \langle(3), (4,5,6,7)\rangle$. То же, что отмечено выше для 4-й строки таблицы смежности, можно отнести и конфигурациям строк с 5-й по 9-ю.

Сформулируем утверждение для рассматриваемого случая, не обобщая его на произвольные случаи. Доказательство этого утверждения будет получено в процессе построения.

***Утверждение* 1:** *Если при построении предельно (3,2)-компактного графа строка $v_i$ таблицы смежностей имеет две вакансии смежности и четыре претендующие на них "черные" вершины с попарно равными конфигурациями, то каждая из вакансий будет заполнена вершинами-претендентами с отличающимися конфигурациями, а выбор варианта между вершинами с равными конфигурациями может быть произвольным.*

Согласно утверждению 1 обе вакансии смежности вершины 4 могут быть заполнены вершиной из (6,7) и вершиной из (8,9). Свобода выбора варианта заполнения вакансии следует из того, что вершины 6 и 7, как и вершины 8 и 9, — висячие и порождены одной и той же вершиной: в первом случае вершиной 2, во втором — вершиной 3. Следовательно, перенумерация вершин 6 в 7 или 8 в 9 никак не изменит рассматриваемый в данный момент граф $G'(V,E')$. Распространив сказанное на вершины 5-9, получим: вершина 5 смежна вершине из (6,7) и вершине из (8,9), вершина 6 — вершинам 4 или 5 и 8 или 9, …, вершина 9 будет смежна вершинам 4 или 5 и 6 или 7. Заметим, что все эти высказывания относительно вариантов заполнения вакансий смежности с 4-й по 9-ю вершин согласуются между собой и непротиворечивы.

Выбираем для определенности соединение ребром вершин 4 и 6 и модифицируем проекции $P'(v_i)| 0 < v_i \leq 9$, множества $\overline{M}^*(v_i)$ «серых» и $M^*(v_i)$ «черных» вершин, а новую таблицу смежности, дополненную ребром 4-6, поместим ниже обновленных проекций и, если этого потребует какая-либо из проекций, будем вносить в нее соответствующие изменения.

$$0\begin{pmatrix}1^{(4,5)},\\2^{(6,7)},\\3^{(8,9)}.\end{pmatrix}\begin{pmatrix}\equiv\equiv\\\equiv\equiv\\\equiv\equiv\end{pmatrix},$$
0,1,2,3,4,5,6,7,8,9;

$$1\begin{pmatrix}0^{(2,3)},\\4^{(6,*)},\\5^{(*,*)}.\end{pmatrix}\begin{pmatrix}\equiv\equiv\\\equiv 0-6\\\equiv 0-6\end{pmatrix},$$
0,1,2,3,4,5,~~6,~~7,8,9;

$$2\binom{0^{(1,3)},}{6^{(4,*)},\atop 7^{(*,*)}.}\binom{\equiv^=_{\overline{0-7}}}{\equiv^{\phantom{0}}_{\overline{0-4,6,7}}},$$
0,1,2,3,~~4~~,5,6,7,8,9;

$$3\binom{0^{(1,2)},}{8^{(*,*)},\atop 9^{(*,*)}.}\binom{\equiv^=_{\overline{0-3,8,9}}}{\equiv^{\phantom{0}}_{\overline{0-3,8,9}}},$$
0,1,2,3,4,5,6,7,8,9;

$$4\binom{1^{(0,5)},}{6^{(2,*)},\atop *^{(*,*)}.}\binom{\equiv^=_{\overline{0-7}}}{\overline{0-7}^{\mathbf{3,7}}},$$
0,1,~~2,3~~,4,5,~~6,7~~,8,9;

$$5\binom{1^{(0,4)},}{*^{(*,*)},\atop *^{(*,*)}.}\binom{\equiv^=_{\overline{0-6}^{0,1,4,5}}}{\overline{0-6}^{0,1,4,5}},$$
0,1,2,3,4,5,6,7,8,9;

$$6\binom{2^{(0,7)},}{4^{(1,*)},\atop *^{(*,*)}.}\binom{\equiv^=_{\overline{0-7}}}{\overline{0-7}^{0-2,\mathbf{3},4,6-9}},$$
0,~~1~~,2,~~3,4~~,5,6,7,8,9;

$$7\binom{2^{(0,6)},}{*^{(*,*)},\atop *^{(*,*)}.}*\binom{\overline{0-4,6,7}^{0,2,6,7}}{\overline{0-4,6,7}^{0,2,6,7}},$$
0,1,2,3,4,5,6,7,8,9;

$$8\binom{3^{(0,9)},}{*^{(*,*)},\atop *^{(*,*)}.}\binom{\overline{0-3,8,9}^{0,3,8,9}}{\overline{0-3,8,9}^{0,3,8,9}},$$
0,1,2,3,4,5,6,7,8,9;

$$9\binom{3^{(0,8)},}{*^{(*,*)},\atop *^{(*,*)}.}\binom{\overline{0-3,8,9}^{0,3,8,9}}{\overline{0-3,8,9}^{0,3,8,9}},$$
0,1,2,3,4,5,6,7,8,9;

**Таблица 3.** Таблица смежностей после добавления ребра 4-6

|       | 0 | 1 | 2 | 3 | 4 | 5 | 6 | 7 | 8 | 9 |
|-------|---|---|---|---|---|---|---|---|---|---|
| **0** | 0 | **1** | **2** | **3** | 4 | 5 | 6 | 7 | 8 | 9 |
| **1** | **0** | 1 | 2 | 3 | **4** | **5** | 6 | 7 | 8 | 9 |
| **2** | **0** | 1 | 2 | 3 | 4 | 5 | **6** | **7** | 8 | 9 |
| **3** | **0** | 1 | 2 | 3 | 4 | 5 | 6 | 7 | **8** | **9** |
| **4** | 0 | **1** | 2 | 3 | 4 | 5 | **6** | ~~7~~ | 8 | 9 |
| **5** | 0 | **1** | 2 | 3 | 4 | 5 | ~~6~~ | 7 | 8 | 9 |
| **6** | 0 | 1 | **2** | 3 | **4** | ~~5~~ | 6 | 7 | 8 | 9 |
| **7** | 0 | 1 | **2** | 3 | **4** | 5 | 6 | 7 | 8 | 9 |
| **8** | 0 | 1 | 2 | **3** | 4 | 5 | 6 | 7 | 8 | 9 |
| **9** | 0 | 1 | 2 | **3** | 4 | 5 | 6 | 7 | 8 | 9 |

Поясним процесс изменений на проекции $P'(1)$:

$$1\binom{0^{(2,3)},}{4^{(*,*)},\atop 5^{(*,*)}.}\binom{\equiv^=_{\overline{0-5}}}{\equiv^{\phantom{0}}_{\overline{0-5}}},$$
0,1,2,3,4,5,6,7,8,9;

====>

$$1\binom{0^{(2,3)},}{4^{(6,*)},\atop 5^{(*,*)}.}\binom{\equiv^=_{\overline{0-6}}}{\equiv^{\phantom{0}}_{\overline{0-6}}},$$
0,1,2,3,4,5,6,7,8,9;

Для удобства ветви проекции будем именовать по их основанию (корню): так в рассматриваемой проекции основание проекции — вершина 1, основания ветвей — 0, 4 и 5, соответствующие им ветви проекции будем называть нулевой, четвертой и пятой. Итак, добавление в исходный граф ребра 4-6 заполняет вакансию второго уровня четвертой ветви вершиной 6, она же исключается из подмножества $M^*(6)$ «черных» вершин, добавляется к подмножеству $\bar{M}^*(6)$ «серых». Кроме того, учитывая, что проекция не должна иметь реплик, вершина 6 добавляется во все значащие (не помеченные символом ≡) множества запрещенных вершин-соискателей вакансий. Аналогично поступаем с остальными проекциями, учитывая при этом изменения таблицы потенциальной смежности: например, в проекции $P'(2)$ мы не просто заполняем вершиной 6 вакансию ее четвертой ветви и, соответственно, корректируем множества $\bar{M}^*_{4,2}(6) = \bar{M}^*_{2,4}(7) := \bar{M}^*_{2,4}(6) \cup 4 = \{0-3,6,7\} \cup 4 = \{0-4,6,7\}$, а сопоставляя эти множества с конфигурациями соответствующих строк таблицы, получим $\bar{M}^*_{2,4}(6) := \{0-4,6,7\} \cup \{V \setminus N'(6)\} = \{0-7\}$, $\bar{M}^*_{2,4}(7)$ при этом остается прежним:

$$2 \begin{pmatrix} 0^{(1,3)}, \\ 6^{(*,*)}, \\ 7^{(*,*)} \end{pmatrix} \begin{pmatrix} \equiv^{\equiv} \\ \equiv^{0-3,6,7} \\ \equiv^{0-3,6,7} \end{pmatrix}, \quad =====> \quad 2 \begin{pmatrix} 0^{(1,3)}, \\ 6^{(4,*)}, \\ 7^{(*,*)} \end{pmatrix} \begin{pmatrix} \equiv^{\equiv} \\ \equiv^{0-7} \\ \equiv^{0-4,6,7} \end{pmatrix}.$$

0,1,2,3,4,5,6,7,8,9;          0,1,2,3,4,5,6,7,8,9;

$$4 \begin{pmatrix} 1^{(0,5)}, \\ *^{(*,*)}, \\ *^{(*,*)} \end{pmatrix} \begin{pmatrix} \equiv^{\equiv} \\ 0-5^{0,1,4,5} \\ 0-5^{0,1,4,5} \end{pmatrix}, \quad ===> \quad 4 \begin{pmatrix} 1^{(0,5)}, \\ 6^{(2,*)}, \\ *^{(*,*)} \end{pmatrix} \begin{pmatrix} \equiv^{\equiv} \\ 0-7 \\ 0-7^{0-2,8,9} \end{pmatrix}, \quad ===> \quad 4 \begin{pmatrix} 1^{(0,5)}, \\ 6^{(2,*)}, \\ *^{(*,*)} \end{pmatrix} \begin{pmatrix} \equiv^{\equiv} \\ 0-7 \\ 0-7^{\mathbf{3,7}} \end{pmatrix}.$$

0,1,2,3,4,5,6,7,8,9;     0,1,2,3,4,5,6,7,8,9;     0,1,2,3,4,5,6,7,8,9;

На примере проекции $P'(4)$ показан еще один из используемых приемов: для второго уровня *-й ветви (с неизвестным основанием, помеченным символом *) множество $\bar{M}^*_{2,*}(4)$ вершин, исключенных из числа претендующих на вакансии второго уровня этой ветви с обычным начертанием, заменено достоверно заполняющими обе (2,*)-вакансии «жирными» вершинами **3** и **7** — в соответствии с отмеченными выше свойствами проекций искомого графа все 10 вершин в любой его 2-уровневой проекции обязательно должны быть упомянуты, тогда как эти вершины (третья и седьмая) внесены в перечень запрещенный всех остальных ($l \neq 2,*$)-вакансий. Поступив аналогично с оставшимися проекциями $P'(v_i) | \ 4 < v_i \leq 9$ подмножествами $M^*(v_i)$, $\bar{M}^*(v_i)$ и таблицей 3 потенциальной смежности, перейдем к следующему шагу, где, используя полученные результаты, произведем выбор варианта смежности.

Из таблицы 3 видно, что в результате выполнения шага 1 число претендующих на вакансии вершин сократилось с $C_{0.2} = 14$ до $C_1 = 9$.

*Шаг 2. Выбор и добавление ребра 5-7*

В пояснениях к утверждению 1 мы уже указывали, что вершина 5 может быть смежной только вершинам 6 или 7 и вершинам 8 или 9. Из полученной на шаге 2 таблицы 3 видим, что вариант соединения ребром вершин 5 и 6 теперь запрещен, поэтому в соответствии с утверждением 1 одна из двух вакансий смежности вершины 5 может быть заполнена только вершиной 7.

Ниже приведены проекции и таблица смежности соответствующие введению в граф $G'(V,E')$ ребра 5-7: $E' := E' + (5\text{-}7)$.

$$0 \begin{pmatrix} 1^{(4,5)}, \\ 2^{(6,7)}, \\ 3^{(8,9)}. \end{pmatrix} \begin{pmatrix} \equiv^= \\ \equiv^= \\ \equiv^= \end{pmatrix},$$
0,1,2,3,4,5,6,7,8,9;

$$1 \begin{pmatrix} 0^{(2,3)}, \\ 4^{(6,*)}, \\ 5^{(7,*)}. \end{pmatrix} \begin{pmatrix} \equiv^= \\ \equiv^{0-7} \\ \equiv^{0-7} \end{pmatrix},$$
0,1,2,3,4,5,6,~~7~~,8,9;

$$2 \begin{pmatrix} 0^{(1,3)}, \\ 6^{(4,*)}, \\ 7^{(5,*)}. \end{pmatrix} \begin{pmatrix} \equiv^= \\ \equiv^{0-7} \\ \equiv^{0-7} \end{pmatrix},$$
0,1,2,3,4,~~5~~,6,7,8,9;

$$3 \begin{pmatrix} 0^{(1,2)}, \\ 8^{(*,*)}, \\ 9^{(*,*)}. \end{pmatrix} \begin{pmatrix} \equiv^= \\ \equiv^{0-3,8,9} \\ \equiv^{0-3,8,9} \end{pmatrix},$$
0,1,2,3,4,5,6,7,8,9;

$$4 \begin{pmatrix} 1^{(0,5)}, \\ 6^{(2,*)}, \\ *^{(*,*)}. \end{pmatrix} \begin{pmatrix} \equiv^= \\ \equiv^{0-7} \\ 0-7^{\mathbf{3,7}} \end{pmatrix},$$
0,1,2,3,4,5,6,7,8,9;

$$5 \begin{pmatrix} 1^{(0,4)}, \\ 7^{(2,*)}, \\ *^{(*,*)}. \end{pmatrix} \begin{pmatrix} \equiv^= \\ \equiv^{0-7} \\ 0-7^{\mathbf{3,6}} \end{pmatrix},$$
0,1,~~2,3~~,4,5,~~6,7~~,8,9;

$$6 \begin{pmatrix} 2^{(0,7)}, \\ 4^{(1,*)}, \\ *^{(*,*)}. \end{pmatrix} \begin{pmatrix} \equiv^= \\ \equiv^{0-7} \\ 0-7^{\mathbf{3,5}} \end{pmatrix},$$
0,1,2,3,4,~~5~~,6,7,8,9;

$$7 \begin{pmatrix} 2^{(0,6)}, \\ 5^{(1,*)}, \\ *^{(*,*)}. \end{pmatrix} * \begin{pmatrix} \equiv^= \\ \equiv^{0-7} \\ 0-7^{\mathbf{3,4}} \end{pmatrix},$$
0,~~1~~,2,~~3,4,5~~,6,7,8,9;

$$8 \begin{pmatrix} 3^{(0,9)}, \\ *^{(*,*)}, \\ *^{(*,*)}. \end{pmatrix} : \begin{pmatrix} \equiv^= \\ 0-3,8,9^{0,3,8,9} \\ 0-3,8,9^{0,3,8,9} \end{pmatrix},$$
0,1,2,3,4,5,6,7,8,9;

$$9 \begin{pmatrix} 3^{(0,8)}, \\ *^{(*,*)}, \\ *^{(*,*)}. \end{pmatrix} \begin{pmatrix} \equiv^= \\ 0-3,8,9^{0,3,8,9} \\ 0-3,8,9^{0,3,8,9} \end{pmatrix},$$
0,1,2,3,4,5,6,7,8,9;

**Таблица 4.** Таблица смежности с ребром 5-7

| | 0 | 1 | 2 | 3 | 4 | 5 | 6 | 7 | 8 | 9 |
|---|---|---|---|---|---|---|---|---|---|---|
| **0** | 0 | **1** | **2** | **3** | 4 | 5 | 6 | 7 | 8 | 9 |
| **1** | **0** | 1 | 2 | 3 | **4** | **5** | 6 | 7 | 8 | 9 |
| **2** | **0** | 1 | 2 | 3 | 4 | 5 | **6** | **7** | 8 | 9 |
| **3** | **0** | 1 | 2 | 3 | 4 | 5 | 6 | 7 | **8** | **9** |
| **4** | 0 | **1** | 2 | 3 | 4 | 5 | **6** | 7 | 8 | 9 |
| **5** | 0 | **1** | 2 | 3 | 4 | 5 | 6 | **7** | 8 | 9 |
| **6** | 0 | 1 | **2** | 3 | **4** | 5 | 6 | 7 | 8 | 9 |
| **7** | 0 | 1 | **2** | 3 | 4 | **5** | 6 | 7 | 8 | 9 |
| **8** | 0 | 1 | 2 | **3** | 4 | 5 | 6 | 7 | 8 | 9 |
| **9** | 0 | 1 | 2 | **3** | 4 | 5 | 6 | 7 | 8 | 9 |

Шаг 2 сократил число претендующих на вакансии вершин с $C_1 = 9$ до $C_2 = 8$.

*Шаг 3. Выбор и добавление ребра 4-8*

Учитывая утверждение 1 в применении его к вершине (строке таблицы) 4, выбираем соединение ребром вершин 4 и 8 и модифицируем таблицу, проекции и подмножества. В таблице закрываем единственную свободную вакансию вершины 4 (строки 4 таблицы) вершиной 8 и скрываем серым шрифтом «черные» вершины, убирая их из числа соискателей вакансии смежности с вершиной 4 (в данном случае убираем вершину 9). Соответственно, вершиной 4 закрываем одну из двух вакансий вершины 8 и скрываем вершину 4 в строке 9.

$0 \begin{pmatrix} 1^{(4,5)}, \\ 2^{(6,7)}, \\ 3^{(8,9)}. \end{pmatrix} \begin{pmatrix} \equiv^= \\ \equiv^= \\ \equiv^= \end{pmatrix}$,

0,1,2,3,4,5,6,7,8,9;

$1 \begin{pmatrix} 0^{(2,3)}, \\ 4^{(6,8)}, \\ 5^{(7,*)}. \end{pmatrix} \begin{pmatrix} \equiv^= \\ \equiv^= \\ \equiv \underline{\mathbf{9}} \end{pmatrix}$,

0,1,2,3,4,5,6,7,8,9;

$2 \begin{pmatrix} 0^{(1,3)}, \\ 6^{(4,*)}, \\ 7^{(5,*)}. \end{pmatrix} \begin{pmatrix} \equiv^{0-7} \\ \equiv^{0-7} \end{pmatrix}$,

0,1,2,3,4,5,6,7,8,9;

$3 \begin{pmatrix} 0^{(1,2)}, \\ 8^{(4,*)}, \\ 9^{(5,*)}. \end{pmatrix} \begin{pmatrix} \equiv^= \\ \equiv \underline{0-5,8,9} \\ \equiv \underline{0-5,8,9} \end{pmatrix}$,

0,1,2,3,~~4,5,~~6,7,8,9;

$4 \begin{pmatrix} 1^{(0,5)}, \\ 6^{(2,*)}, \\ 8^{(3,7)}. \end{pmatrix} \begin{pmatrix} \underline{\mathbf{9}} \\ \equiv^= \\ \equiv^= \end{pmatrix}$,

0,1,2,3,4,5,6,7,~~8,9~~;

$5 \begin{pmatrix} 1^{(0,4)}, \\ 7^{(2,8)}, \\ 9^{(3,6)}. \end{pmatrix} \begin{pmatrix} \equiv^= \\ \equiv^= \\ \equiv^= \end{pmatrix}$,

0,1,2,3,4,5,6,7,~~8,9~~;

$$6\begin{pmatrix}2^{(0,7)},\\4^{(1,8)},\\9^{(3,5)}.\end{pmatrix}\begin{pmatrix}\equiv\\\equiv\\\equiv\end{pmatrix},$$

0,1,2,3,4,5,6,7,~~8,9~~;

$$8\begin{pmatrix}3^{(0,9)},\\4^{(1,6)},\\7^{(2,5)}.\end{pmatrix}\begin{pmatrix}\equiv\\\equiv\\\equiv\end{pmatrix},$$

0,~~1,2~~,3,~~4,5,6,7~~,8,9;

$$7\begin{pmatrix}2^{(0,6)},\\5^{(1,9)},\\8^{(3,4)}.\end{pmatrix}\begin{pmatrix}\equiv\\\equiv\\\equiv\end{pmatrix},$$

0,1,2,3,4,5,6,7,~~8,9~~;

$$9\begin{pmatrix}3^{(0,8)},\\5^{(1,7)},\\6^{(2,4)}.\end{pmatrix}\begin{pmatrix}\equiv\\\equiv\\\equiv\end{pmatrix},$$

0,~~1,2~~,3,~~4,5,6,7~~,8,9;

**Таблица 5**. Таблица смежностей с ребром 4-8

|   | 0 | 1 | 2 | 3 | 4 | 5 | 6 | 7 | 8 | 9 |
|---|---|---|---|---|---|---|---|---|---|---|
| **0** | 0 | **1** | **2** | **3** | 4 | 5 | 6 | 7 | 8 | 9 |
| **1** | **0** | 1 | 2 | 3 | **4** | **5** | 6 | 7 | 8 | 9 |
| **2** | **0** | 1 | 2 | 3 | 4 | 5 | **6** | **7** | 8 | 9 |
| **3** | **0** | 1 | 2 | 3 | 4 | 5 | 6 | 7 | **8** | **9** |
| **4** | 0 | **1** | 2 | 3 | 4 | 5 | **6** | 7 | **8** | ~~9~~ |
| **5** | 0 | **1** | 2 | 3 | 4 | 5 | 6 | **7** | ~~8~~ | **9** |
| **6** | 0 | 1 | **2** | 3 | **4** | 5 | 6 | 7 | ~~8~~ | **9** |
| **7** | 0 | 1 | **2** | 3 | 4 | **5** | 6 | 7 | **8** | ~~9~~ |
| **8** | 0 | 1 | 2 | **3** | **4** | ~~5~~ | ~~6~~ | **7** | 8 | 9 |
| **9** | 0 | 1 | 2 | **3** | ~~4~~ | **5** | **6** | ~~7~~ | 8 | 9 |

Модифицируя $P'(1)$, получаем ребро **5-9**, заполняем последнюю вакансию в строке 5 таблицы 5. Так как все вакансии в строке 5 теперь заполнены $f(5) = 3$, $v(5) = 0$, то $b(5) = 0$ избыточную "чёрную" вершину 8 в этой строке зачёркиваем: ребра 5-8 в нашем графе существовать не может. Производим те же действия в строках 5, 9 и 8. Продолжая модификацию проекций, дадим дополнительные пояснения к проекции $P'(4)$. Заполнив в ней свободную вакансию первого уровня вершиной **8**, обнаруживаем, что второй уровень этой ветви уже заполнен вершинами 3 и **7**, что указывает на смежность вершины 8 с этими вершинами, — внесём соответствующее изменение в таблицу смежности искомого графа, оно коснётся лишь вершины 7, так как смежность 8 с 3 уже была задана таблицей. Все три вакансии восьмой строки окажутся при этом заполненными, оставшуюся в строке 8 "чёрную" вершину 6 извлекаем из числа потенциально смежных, зачеркнув её, и делаем то же самое с вершиной 8 в строке 6. В строке **6** при этом остаётся всего одна «чёрная» вершина — **9**, претендующая на единственную в этой строке вакансию (из необходимых трёх ($s = 3$) в строке 6 всего две "жирные" вершины).

Заполним эту вакансию вершиной **9** (меняем ее шрифт на полужирный), добавим в выстраиваемый граф ребро (6-9), строка 9 при этом также дополняется «жирной» вершиной **6**.

Таким образом, введя на этом шаге ребро 4-8, мы однозначно получили все три недостающих в искомом графе ребра (ребра 5-9, 8-7 и 6-9), таблица смежностей потеряла при этом статус потенциальной и приобрела окончательный вид таблицы смежностей искомого графа

Общеизвестно, что отсечение гарантированно неперспективных множеств вариантов является базовой идеей всех точных алгоритмов [33]. Выше было показано, что число вакансий смежности на каждом шаге сокращается именно за счет отсечения гарантированно неприемлемых (противоречащих заданным свойствам искомого графа) вариантов, поэтому основанное на этом решение системы проекций является точным.

Продолжив модификацию оставшихся проекций с учетом полученной на этом шаге таблицы 5, убедимся в том, что все двухуровневые проекции вершинно полны (все вершины из $V$, помещенные под каждой проекцией, упомянуты в ней (стали серыми), и полученный таким образом граф полностью соответствует заданным условиям: все 10 двухуровневых проекций являются вершинно полными (содержат в себе все поименованные в графе вершины 10 вершин), двухуровневость всех проекций графа указывает на его соответствие заданному диаметру $d = 2$. Проекции $P'(2)$ и $P'(3)$, предшествующие полученному в проекции $P'(4)$ решению, для большей наглядности оставлены в том же виде, в каком они были до рассмотрения $P'(4)$.

Итак, первый шаг (добавление в граф ребра 4-6) уменьшил число недостающих ребер до пяти, второй шаг сократил число недостающих ребер до четырех, а уже третий шаг (добавление ребра 4-8) однозначно определил оставшиеся четыре ребра. Обратим внимание на то, что выбор ребер на первом и втором шагах был основан на утверждении 1, и для первого шага мы могли бы выбрать любое другое ребро (4-7, 4-8, 4-9, или 5-6, 5-7, 5-8, 5-9, или же 6-8, 6-9 и т. д. в соответствии с упомянутой выше таблицей), вакансии для выбора ребра на втором шаге изменятся соответственно… Тогда были бы получены графы с другой нумерацией вершин, но изоморфные полученному выше, в чем нетрудно убедиться выполнив подобные подстановки.

Таким образом, каждый связанный с выбором ребра шаг уменьшает число искомых ребер минимум на единицу, уменьшение при этом числа возможных реберных комбинаций приведено для каждого шага. Асимптотическая временная сложность $T_{step}$ одного шага пропорциональна числу проекций *n*, поэтому . Учитывая, что выбор ребер на каждом шаге обоснован заданными свойствами графа (проекций графа), можно предположить, что для предельно компактных графов число $n_{step}$ шагов всегда будет меньшим числа неизвестных $n_x$ ребер: $n_{step} < (n_x = ns/2 - (n-1))$, или $n_{step} = o(n(s-2))$. Перемножив эти величины получим общую временную сложность *T* поиска предельно компактных графов: $T_{step} = O(n) \cdot o(n(s-2)) = o(n^2(s-2))$. А так как число линков *s* в вычислительных системах имеет физические ограничения, величину (*s* – 2) в сравнении с увеличивающимся *n* можно считать постоянной, и асимптотическая сложность предложенного в применении к таким системам метода является квадратической: $T_{step} = o(n^2)$.

Итак, решив систему проекций исходного графа, представленного в данном случае каркасом искомого графа, мы получили все недостающие для его построения ребра. Каждая из результирующих 2-уровневых проекций хотя и является вершинно полной, но не дает описания всех ребер полученного графа. Для описания графа единственной проекцией, достаточно надстроить любую из этих проекций еще одним уровнем, используя для этого результирующую таблицу 5, либо взяв необходимую информацию о смежности вершин второго уровня из других проекций. Покажем это примерами полных проекций *P*(0), *P*(3) и *P*(8); для построения полученного графа достаточно любой из них:

$$0 \begin{pmatrix} 1^{(4,8)}, 5^{(7,9)}, \\ 2^{(6,4,9)}, 7^{(5,8)}, \\ 3^{(8,4,7)}, 9^{(5,6)}. \end{pmatrix}, \quad 3 \begin{pmatrix} 0^{(1,4,5)}, 2^{(6,7)}, \\ 8^{(4,1,6)}, 7^{(2,5)}, \\ 9^{(5,1,7)}, 6^{(2,4)}. \end{pmatrix}, \quad 8 \begin{pmatrix} 3^{(0,1,2)}, 9^{(5,6)}, \\ 4^{(1,0,5)}, 6^{(2,9)}, \\ 7^{(2,0,6)}, 5^{(1,9)}. \end{pmatrix}.$$

Из приведенных выше надстроенных третьим уровнем проекций можно увидеть, что реплики, впервые появившиеся при этом на уровне 3, имеют оригиналами только вершины уровня 2. Следовательно, длина минимального цикла в каждой из проекций, определяемая суммой номеров уровней оригинальной вершины и ее реплики, равна пяти. Аналогичную картину нетрудно увидеть и из остальных проекций, т. е. обхват нашего 10-вершинного степени 3 графа равен пяти, и этот граф не что иное как всем известная (3,5)- клетка, или граф Петерсена [34], он же, как обладающий при этих значениях степени и обхвата минимальным порядком, — граф Мура. Изоморфизм полученного нами графа графам, поименнованным выше, следует из приведенного ниже рисунка, в котором на общеизвестные изображения последних нанесена нумерация первого.

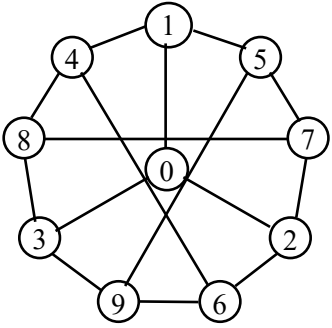 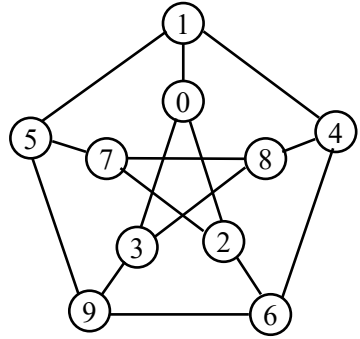

Приведенный здесь пример генерации предельно (3,2)-компактного графа завершается успешно: система проекций сходится. Однако известно, что такие графы для произвольных сочетаний степени $s$ и диаметра $d$ довольно редки и известны лишь для небольших их значений [35]. Поэтому, за редкими исключениями, исходная для построения предельно компактного графа система проекций с заданными значениями $s$ и $d$ не имеет решения. В таком случае число "черных", претендующих на вакансии, вершин в одной или в нескольких строках промежуточной таблицы потенциальной смежности станет меньшим числа вакансий смежности этих строк. Это укажет на несовместность системы проекций при заданных начальных условиях и, соответственно, на невозможность существования удовлетворяющего этим условиям графа. К счастью, подобная, связанная с требованием максимального обхвата, эксклюзивность не свойственна компактным графам с промежуточными числами вершин из формулы 1, и таких графов достаточно много. Ниже, а также в приложениях П1 и П2 в качестве примеров приведены проекции полученных нами 30(3,4)– и 15(4,2)–компактных графов число вершин в которых меньше, чем полученные из (1) предельные значения: $n = 46$ и $n = 17$, соответственно.

При построении $n(s,d)$-компактного графа с числом вершин $n$, меньшим предельного, $d$-уровневые проекции искомого графа в отличие от рассмотренного выше примера, где $c = 0$, содержат реплики (в этом случае $c = 1 + s\sum_{i=1}^{d}(s-1)^{i-1} - n > 0$), что и должно быть учтено свойствами $Q(P'(G))$. Дабы не перегружать статью, ограничимся приведением здесь лишь исходных и полученных из них полных проекций графов с требуемыми свойствами.

Ниже приведены: слева — исходная вершинно полная 4-уровневая проекция каркаса генерируемого 30(3,4)-компактного графа ($n = 30$, $s = 3$,

$d = 4$) и справа — также 4-уровневая, но уже полная проекция искомого графа, полученная в результате решения системы из тридцати исходных проекций, описывающих остовное дерево. К начальным условиям, обусловливающим корректирование проекций, здесь добавлено условие обхвата: длина наименьшего цикла в искомом графе равна восьми — $g(G) = 8$. Так как в проекции сумма номеров уровней вершины и ее реплик не превышает длины содержащего эту вершину цикла, то учитывая, что число уровней в проекциях равно четырем ($d = 4$), и вершина, и ее реплика в каждой проекции могут находиться только на четвертом уровне. Это означает также, что и вершина, и ее реплики не должны принадлежать одной и той же ветви проекции, иначе длина цикла и, соответственно, обхват будут меньше заданного $g(G) = 8$. Добавление условия обхвата существенно ограничило подмножества вершин, претендующих на ту или иную вакансию смежности, облегчило выбор подстановочных ребер и уменьшило число шагов. Нетрудно убедиться, что полученный граф является (3,8)-клеткой: обхват его равен восьми, а число вершин — нижнему пределу для четных обхватов, в связи с чем он может быть отнесен к графам Мура [36].

$$0\begin{pmatrix} 1^{(4^{(10^{(*,*)}},11^{(*,*)},5^{(12^{(*,*)},13^{(*,*)})})} \\ 2^{(6^{(14^{(22,23)},15^{(24,25)},7^{(16^{(26,27)},17^{(28,29)})})} \\ 3^{(8^{(18^{(*,*)},19^{(*,*)},9^{(20^{(*,*)},21^{(*,*)})})} \end{pmatrix} \to 0\begin{pmatrix} 1^{(4^{(10^{(22,28)},11^{(25,27)},5^{(12^{(23,29)},13^{(24,26)})})} \\ 2^{(6^{(14^{(22,23)},15^{(24,25)},7^{(16^{(26,27)},17^{(28,29)})})} \\ 3^{(8^{(18^{(23,27)},19^{(24,28)},9^{(20^{(22,26)},21^{(25,29)})})} \end{pmatrix}$$

Приведем теперь результат использования предложенного метода для получения 15(4,2)-компактного графа из исходного графа, отличного от остовных деревьев, примененных в рассмотренных выше примерах. Предположим, нам требуется построить ВС с централизованным управлением, устойчивым к отказам кратности два. Для создания такого отказоустойчивого центрального ядра, исходный граф здесь образован 3-циклом с вершинами 0, 1, 2; затем к каждой вершине этого цикла добавим по две смежные вершины (соответственно, 3 и 4, 5 и 6, 7 и 8), задав отношения смежности для каждой из них с вершинами, оставшимися неупомянутыми: 3-9, 4-10, …, 8-14.

$$0\begin{matrix} 1^{(2^{(0,7,8)},5^{(*,11,*)},6^{(*,12,*)})}, \\ 2^{(1^{(0,5,6)},7^{(*,*,13)},8^{(*,*,14)})}, \\ 3^{(9^{(*,*,*)},12^{(6,*,*)},14^{(*,8,*)})}, \\ 4^{(10^{(*,*,*)},11^{(5,*,*)},13^{(*,7,*)})} \end{matrix} \to 0\begin{matrix} 1^{(2^{(0,7,8)},5^{(10,11,14)},6^{(9,12,13)})}, \\ 2^{(1^{(0,5,6)},7^{(9,10,13)},8^{(11,12,14)})}, \\ 3^{(9^{(6,7,11)},12^{(6,8,10)},14^{(5,8,13)})}, \\ 4^{(10^{(5,7,12)},11^{(5,8,9)},13^{(6,7,14)})} \end{matrix}.$$

В достоверности полученных решений нетрудно убедиться, построив из данных здесь полных проекций 30(3,4)-компактного графа и 15(4,2)-

графа с 3-циклом, систему 4-уровневых вершинно полных проекций первого и систему 3-уровневых вершинно полных проекций второго — (приложения П1 и П2, соответственно). Наличие всех помеченных вершин в каждой из проекций указывает на то, что эксцентриситеты любой из вершин, а следовательно, и диаметры графов не превышают 4-х и 3-х, соответственно, что и было задано начальными условиями для их построения.

### 3. Заключение

В работе впервые описаны формальные процедуры получения графов с заданными свойствами. Основой метода являются оригинальное их описание с помощью проекций, свойства которых биективны свойствам искомого графа. Решение системы проекций в отношении этих свойств позволяет либо получить проективное описание такого графа, либо, при выявлении ее несовместности, доказать невозможность его существования. Биекция свойств топологий интерконнекта вычислительных систем и свойств генерируемых графов позволяет использовать предлагаемую методику в заполнении ниши теоретических основ *детерминированного* проектирования и использования крупномасштабных вычислительных систем и суперкомпьютеров. Именно это, прежде всего, и явилось предпосылкой разработки как оригинального метода проективного описания графов, так и его последующего развития и использования, включая предлагаемую в данной работе методику.

Проблема построения графов с заданными свойствами является чрезвычайно насущной в теории графов. Еще в 1981 году Вонг в статье [37] отметил: "Регулярный граф с заданными обхватом и валентностью получить довольно сложно. Особенно, когда обхват или валентность графа велики. И только с помощью компьютера, *некоторые* из таких графов могут быть успешно получены". С тех пор в отношении формальных методов синтеза графов с заданными свойствами практически мало что изменилось. Автор надеется, что использование предложенного метода не только восполнит эти пробелы в теории графов, но позволит производить менее трудоемкий и более качественный систематический анализ и сопоставление исследуемых графов. Естественным образом это скажется также и на всех приложениях, использующих аппарат теории графов.

Латышеву, благодаря которым имею возможность продолжать исследования по этой тематике.

## СПИСОК ЛИТЕРАТУРЫ

1. *Задорожный А.Ф., Мелентьев В.А.* О совместимости топологий параллельных задач и систем // 13-я мультиконференция по проблемам управления, материалы конференции «Информационные технологии в управлении» (ИТУ-2020). – 2020. – СПб.: АО «Концерн «ЦНИИ «Электроприбор». С. 162–164.
2. *Каляев И.А., Левин И.И., Семерников Е.А.* Реконфигурируемые вычислительные системы. //Гироскопия и навигация. - М.: Изд-во ЦНИИ "Электроприбор", 2009. http://fpga.parallel.ru/papers/kaljaev4.pdf
3. *Каравай М. Ф., Подлазов В. С.* Оптимальные отказоустойчивые многомерные торы на основе малопортовых маршрутизаторов и хабов //Проблемы управления. – 2020. – Т. 5. – №. 0. – С. 56-64.
4. *Мелентьев В. А.* Компактные структуры вычислительных систем и их синтез //Управление большими системами: сборник трудов. – 2011. – №. 32. – С. 241–261.
5. *Раппопорт А. М.* Метрические характеристики графов сетей коммуникаций //Труды Института системного анализа Российской академии наук. – 2005. – Т. 14. – С. 141-147.
6. *Мелентьев В. А.* О топологической масштабируемости вычислительных систем //Управление большими системами: сборник трудов. – 2015. – №. 58. – С.115–143.
7. *Erdős P., Rényi A.* On Random Graphs. I // Publicationes Mathematicae. — 1959. — Vol. 6. — P. 290-297.
8. *Gilbert E.N.* Random graphs. //The Annals of Mathematical Statistics. — 1959. — Vol. 30, N4. — P. 1141-1144.
9. *De Vries J.* Over vlakke configuraties waarin elk punt met twee lijnen incident is. Verslagen en Mededeelingen der Koninklijke Akademie voor Wetenschappen, Afdeeling Natuurkunde (3). 1889. — Vol. 6. — P. 382-407.
10. *Brinkmann G, Goedgebeur J, Van Cleemput N.* The history of the generation of cubic graphs. J Chem Inf Model. 2013 Apr 1;5:67-89.
11. *Balaban AT.* Valence-isomerism of cyclopolyenes. Revue Roumaine de chimie. 1966 Jan 1;11(9):1097.
12. *Bussemaker FC, Cobeljic S, Cvetkovic DM, Seidel JJ.* Computer investigation of cubic graphs. EUT report. WSK, Dept. of Mathematics and Computing Science. 1976;76.

13. *Фараджев И. А.* Конструктивное перечисление однородных графов, УМН, 31:1(187) (1976), 246.
14. *McKay BD, Royle GF.* Constructing the cubic graphs on up to 20 vertices. Department of Mathematics, University of Western Australia; 1985.
15. *Brinkmann G.* Generating cubic graphs faster than isomorphism checking. Universität Bielefeld. SFB 343. Diskrete Strukturen in der Mathematik; 1992.
16. *Meringer M.* Fast generation of regular graphs and construction of cages //Journal of Graph Theory. – 1999. – Vol. 30. – N 2. – P. 137-146.
17. *Brinkmann G., Goedgebeur J., McKay B. D.* Generation of cubic graphs //Discrete Mathematics and Theoretical Computer Science. – 2011. – Vol. 13. – N 2. – P. 69--79.
18. *Balaban A. T., Banciu M.* Schemes and transformations in the (CH) 2k series: Valence isomers of [8]-and [10] annulene //Journal of Chemical Education. – 1984. – Vol. 61. – N 9. – P. 766.
19. *Bretto A., Gillibert L.* G-graphs for the cage problem: a new upper bound //Proceedings of the 2007 international symposium on Symbolic and algebraic computation. – 2007. – C. 49-53.
20. Cage Graph — from Wolfram MathWorld. https://mathworld.wolfram.com/CageGraph.html
21. *Мелентьев В. А.* Скобочная форма описания графов и ее использование в структурных исследованиях живучих вычислительных систем //Автометрия. – 2000. – N 4. – С. 36–51.
22. *Мелентьев, В. А.* Формальный подход к исследованию структур вычислительных систем. //Вестник Томского госуниверситета. – 2005. Приложение 14. – С. 167–172.
23. *Мелентьев, В. А.* Аналитический подход к синтезу регулярных графов с заданными значениями порядка, степени и обхвата //Прикл. дискр. мат. – 2010. –Т. 2. N 8 – С. 74–86.
24. *Корнюшко, В. Ф., Панов, А. В., Богунова, И. В., Николаева, О. М., & Флид, А. А.* Системный подход к информационной поддержке фармацевтической разработки готовых лекарственных средств //Тонкие химические технологии. – 2018. – Т. 13. – №. 2. – С. 91–99
25. *Елисеев А. И., Минин Ю. В., Мартемьянов Ю. Ф.* К вопросу об исследовании графов сетевых информационных систем на толерантность в отношении достижимости вершин //Вестник Воронежского института ФСИН России. – 2011. – №. 2. – С. 55–57.
26. *Volkova A.* A Technical Translation of Melentiev's Graph Representation Method with Commentary. – 2018. – University Honors Theses. Paper 503. DOI: 10.15760/honors.507.

27. *Мелентьев В. А.* Проблемы изоморфизма и толерантности графов в теории отказоустойчивости систем //Труды IV Международной конференции «Идентификация систем и задачи управления» SICPRO'05. Москва, ИПУ РАН. – 2005. – С. 28–30.
28. *Мелентьев В. А.* Использование метода Мелентьева представления графов для выявления клик и анализа топологий вычислительных систем //Theoretical & Applied Science. – 2018. –12(68). – С. 201–211.
29. *Мелентьев В. А.* Использование метода Мелентьева представления графов для выявления и перечисления циклов заданной длины // Theoretical & Applied Science. – 2018. –11(67). – С. 85–91.
30. *Melent'ev V.A.* Author's approach to topological modeling of parallel computation systems. // Journal of Mechanics of Continua and Mathematical Sciences – 2020 – Special Issue. 8. – P. 224–237.
31. *Мелентьев В. А.* Компактные графы и детерминированный алгоритм их синтеза // Прикл. дискр. мат. – 2011. – Приложение 4. – С. 94–96.
32. *Мелентьев В. А.* Метрика, цикломатика и синтез топологии систем и сетей связи // Труды шестой Международной конференции «Параллельные вычисления и задачи управления PACO'2012, Москва, 24–26 октября 2012». – 2012. – С. 10–25.
33. *Губко М.В.* Алгоритмические методы решения задач дискретной оптимизации. https://docplayer.ru/41920534-Lekciya-z-algoritmicheskie-metody-resheniya-zadach-diskretnoy-optimizacii.html.
34. *Харари Ф.* Теория графов. – М.: Мир, 1978. – 300 с.
35. *Wong P.* Cages — a survey // Journal of Graph Theory. – 1982. – T. 6. – N 1. – C. 1–22.
36. *Dégila J. R., Sanso B.* A survey of topologies and performance measures for large-scale networks // IEEE Communications Surveys & Tutorials. – 2004. – T. 6. – № 4. – C. 18–31.
37. *Wong P. K.* On the smallest graphs of girth 10 and valency 3 // Discrete Mathematics. – 1983. – T. 43. – № 1. – C. 119–124.

# REFERENCES

1. *Zadorozhny A.F., Melentyev V.A.* About Compatibility of Topologies of Parallel Problems and Systems //13-ya mul'tikonferentsiya po problemam upravleniya, materialy konferentsii «Informatsionnye tekhnologii v upravlenii» (ITU-2020). – 2020. – SPb.: AO «Kontsern «TSNII «ElektropriboR». C. 162–164. (In Russ.).
2. *Kalyayev I.A., Levin I.I., Semernikov Ye.A.* Rekonfiguriruyemye vychislitel'nye sistemy. //Giroskopiya i navigatsiya. - M.: Izd-vo TSNII "Elektropribor", 2009. http://fpga.parallel.ru/papers/kaljaev4.pdf
3. *Karavai M. F., Podlazov V. S.* Optimum multidimensional tori based on low-port routers and hubs //Problemy Upravleniya. – 2020. – T. 5. – C. 56-64. DOI: 10.25728/pu.2020.5.7. (In Russ.).
4. *Melent'ev V.A.* Compact structures of computer systems and their synthesis // Upravlenie bol'shimi systemami. – 2011. – 32. – P. 241–261. (In Russ.).
5. *Rappoport A. M.* Metricheskiye kharakteristiki grafov setey kommunikatsiy //Trudy Instituta sistemnogo analiza Rossiyskoy akademii nauk. – 2005. – T. 14. – S. 141-147. (In Russ.).
6. *Melent'ev V.A.* On topological scalability of computing systems //Upravlenie bol'shimi systemami. – 2015. – Issue 58. – P. 115–143. (In Russ.).
7. *Erdős P., Rényi A.* On Random Graphs. I // Publicationes Mathematicae. — 1959. — Vol. 6. — P. 290-297.
8. *Gilbert E.N.* Random graphs. //The Annals of Mathematical Statistics. — 1959. — Vol. 30, N4. — P. 1141-1144.
9. *De Vries J.* Over vlakke configuraties waarin elk punt met twee lijnen incident is. Verslagen en Mededeelingen der Koninklijke Akademie voor Wetenschappen, Afdeeling Natuurkunde (3). 1889. — Vol. 6. — P. 382-407.
10. *Brinkmann G, Goedgebeur J, Van Cleemput N.* The history of the generation of cubic graphs. J Chem Inf Model. 2013 Apr 1;5:67-89.
11. *Balaban AT.* Valence-isomerism of cyclopolyenes. Revue Roumaine de chimie. 1966 Jan 1;11(9):1097.
12. *Bussemaker FC, Cobeljic S, Cvetkovic DM, Seidel JJ.* Computer investigation of cubic graphs. EUT report. WSK, Dept. of Mathematics and Computing Science. 1976;76.
13. *Faradzhev I. A.* Konstruktivnoye perechisleniye odnorodnykh grafov, UMN, 31:1(187) (1976), 246.
14. *McKay BD, Royle GF.* Constructing the cubic graphs on up to 20 vertices. Department of Mathematics, University of Western Australia; 1985.

15. *Brinkmann G.* Generating cubic graphs faster than isomorphism checking. Universität Bielefeld. SFB 343. Diskrete Strukturen in der Mathematik; 1992.
16. *Meringer M.* Fast generation of regular graphs and construction of cages //Journal of Graph Theory. – 1999. – Vol. 30. – N 2. – P. 137-146.
17. *Brinkmann G., Goedgebeur J., McKay B. D.* Generation of cubic graphs //Discrete Mathematics and Theoretical Computer Science. – 2011. – Vol. 13. – N 2. – P. 69--79.
18. *Balaban A. T., Banciu M.* Schemes and transformations in the (CH) 2k series: Valence isomers of [8]-and [10] annulene //Journal of Chemical Education. – 1984. – Vol. 61. – N 9. – P. 766.
19. *Bretto A., Gillibert L.* G-graphs for the cage problem: a new upper bound //Proceedings of the 2007 international symposium on Symbolic and algebraic computation. – 2007. – C. 49-53.
20. https://mathworld.wolfram.com/CageGraph.html
21. *Melentiev V. A.* The bracket form of graph description and its use in structural investigation of robust computer systems //Optoelectronics Instrumentation and Data Processing. – 2000. – 4. – P. 34–47.
22. *Melent'ev V. A.* Formal'nyy podkhod k issledovaniyu struktur vychislitel'nykh sistem. //Vestnik Tomskogo gosuniversiteta. – 2005. Prilozheniye 14. – S. 167–172. (In Russ.).
23. *Melent'ev V.A.* An analytical approach to the synthesis of regular graphs with preset values of the order, degree and girth //Prikl. Diskr. Mat. –2010. – 2(8). – P. 74–86. (In Russ.).
24. *Kornushko V.F., Panov A.V., Bogunova I.V., Nikolayeva O.M., Flid A.A.* System approach to informational support of pharmaceutical development of finished medicinal products //Fine Chemical Technologies. – 2018. –13(2). – P. 91–99. (In Russ.).
25. *Eliseev A.I., Minin Yu.V., Martemjyanov Ju,F.* Tolerance research of graphs of network information systems in case of the reaching its vertexes // Vestnik of Voronezh Institute of the Russian Federal Penitentiary Service. – 2011. – 2. – P. 55–57. (In Russ.).
26. *Volkova A.* A Technical Translation of Melentiev's Graph Representation Method with Commentary. – 2018. – University Honors Theses. Paper 503. DOI: 10.15760/honors.507.
27. *Melent'ev V.A.* Problems of isomorphism and tolerance of graphs in the systems' fault-tolerance theory //Proceedings of the IV International Conference System Identification and Control Problems SICPRO'05., 25-28 January 2005.: Institute of Control Sciences, Moscow, Russia. – 2005, – P. 532–549. (In Russ.).

28. *Melent'ev V.A.* Use of Melentiev's graph representation method for detection of cliques and the analysis of topologies of computing systems // Theoretical & Applied Science. – 2018. –12(68). – P. 201–211. (In Russ.).
29. *Melent'ev V.A.* Use of Melentiev's graph representation method for identification and enumeration of circuits of the given length // Theoretical & Applied Science. – 2018. – 11(67). – P. 85–91. (In Russ.).
30. *Melent'ev V.A.* Author's approach to topological modeling of parallel computation systems. // Journal of Mechanics of Continua and Mathematical Sciences – 2020 – Special Issue. 8. – P. 224–237.
31. *Melent'ev V.A.* Compact graphs and the deterministic algorithm for their synthesis // Prikl. Diskr. Mat., – 2011. – Supplement 4. – P. 94–96. (In Russ.).
32. *Melent'ev V.A.* The Metric, Cyclomatic and Synthesis of Topology of Systems and Networks // Proceedings of the Sixth International Conference "Parallel Computation and Control Problems" PACO'2012, October 24–26 2012: Moscow, Russia. ICS RAS. – 2012. – P. 10–25. (In Russ.).
33. *Goubko M.V.* Algoritmicheskie-metody-resheniya-zadach-diskretnoy-optimizacii. https://docplayer.ru/41920534-Lekciya-z-algoritmicheskie-metody-resheniya-zadach-diskretnoy-optimizacii.html.
34. Harary F. Graph Theory. – M.: Mir, 1978. – 300 s.
35. *Wong P.* Cages — a survey //Journal of Graph Theory. – 1982. – T. 6. – N 1. – C. 1–22.
36. *Dégila J. R., Sanso B.* A survey of topologies and performance measures for large-scale networks //IEEE Communications Surveys & Tutorials. – 2004. – T. 6. – № 4. – C. 18–31.
37. *Wong P. K.* On the smallest graphs of girth 10 and valency 3 //Discrete Mathematics. – 1983. – T. 43. – № 1. – C. 119–124.

# ПРИЛОЖЕНИЕ П1

Ниже приведены проекции 30(3,4)-компактного графа *G*, в которых **начальные** проекции исходного графа (их вершины выделены жирным шрифтом) совмещены с **итоговыми** проекциями найденного графа, полученными в результате решения системы начальных проекций. Здесь вершины, дополняющие исходный граф до искомого, показаны обычным шрифтом; в каждой отдельно взятой **начальной** проекции эти вершины неизвестны и обозначаются звездочкой — "*".

Число возможных вариантов построения 30-вершинного кубического графа $C_0$ = 27405. Уже одним только построением 30-вершинного каркаса с заданным диаметром *d* = 4 на шаге 0.1 мы добиваемся сокращения числа вариантов с $C_0$ = 27405 до $C_{0.1}$ = 1140, т. е. более чем в 24 раза. Решение начальной системы проекций здесь достигнуто за 11 шагов, в результате чего получены все недостающие 16 ребер:

1-й шаг: добавлено ребро 10-22
2-й шаг: добавлено ребро 20-22
3-й шаг: добавлено ребро 20-26, получено ребро 11-27
4-й шаг: добавлено ребро 11-27
5-й шаг: добавлено ребро 26-13, получено ребро 12-23
6-й шаг: добавлено ребро 12-23
7-й шаг: добавлено ребро 13-24, получены ребра 11-25, 21-25
8-й шаг: добавлены ребра 11-25, 21-25
9-й шаг: добавлено ребро 10-28, получены ребра 12-29, 21-29
10-й шаг: добавлены ребра 12-29, 21-29
11-й шаг: добавлено ребро 19-28, получены недостающие 3 ребра: 18-23, 18-27, 19-24.

Итак, искомый компактный 30(3,4)-граф получен: каждая из тридцати (с 0-й по 29-ю) результирующих 4-уровневых проекций содержит все 30 вершин, что удостоверяет равенство диаметра найденного графа заданному *d* = 4

$$0\begin{Bmatrix} 1^{(4^{(10^{(22,28)}},11^{(25,27)})},5^{(12^{(23,29)}},13^{(24,26)})}), \\ 2^{(6^{(14^{(22,23)}},15^{(24,25)})},7^{(16^{(26,27)}},17^{(28,29)})}), \\ 3^{(8^{(18^{(23,27)}},19^{(24,28)})},9^{(20^{(22,26)}},21^{(25,29)})}). \end{Bmatrix},$$

$$1\begin{Bmatrix} 0^{(2^{(6^{(14,15)}},7^{(16,17)})},3^{(8^{(18,19)}},9^{(20,21)})}), \\ 4^{(10^{(22^{(14,20)}},28^{(17,19)})},11^{(25^{(15,21)}},27^{(16,18)})}), \\ 5^{(12^{(23^{(14,18)}},29^{(17,21)})},13^{(24^{(15,19)}},26^{(16,20)})}). \end{Bmatrix},$$

$$2\begin{Bmatrix} 0^{(1^{(4^{(10,11)}},5^{(12,13)})},3^{(8^{(18,19)}},9^{(20,21)})}), \\ 6^{(14^{(22^{(10,20)}},23^{(12,18)})},15^{(24^{(13,19)}},25^{(11,21)})}), \\ 7^{(16^{(26^{(13,20)}},27^{(11,18)})},17^{(28^{(10,19)}},29^{(12,21)})}). \end{Bmatrix},$$

$$3\begin{Bmatrix} 0^{(1^{(4^{(10,11)}},5^{(12,13)})},2^{(6^{(14,15)}},7^{(16,17)})}), \\ 8^{(18^{(23^{(12,14)}},27^{(11,16)})},19^{(24^{(13,15)}},28^{(10,17)})}), \\ 9^{(20^{(22^{(10,14)}},26^{(13,16)})},21^{(25^{(11,15)}},29^{(12,17)})}). \end{Bmatrix},$$

$$4\begin{Bmatrix} 1^{(0^{(2^{(6,7)}},3^{(8,9)})},5^{(12^{(23,29)}},13^{(24,26)})}), \\ 10^{(22^{(14^{(6,23)}},20^{(9,26)})},28^{(17^{(7,29)}},19^{(8,24)})}), \\ 11^{(25^{(15^{(6,24)}},21^{(9,29)})},27^{(16^{(7,26)}},18^{(8,23)})}). \end{Bmatrix},$$

$$5\begin{Bmatrix} 1^{(0^{(2^{(6,7)}},3^{(8,9)})},4^{(10^{(22,28)}},11^{(25,27)})}), \\ 12^{(23^{(14^{(6,22)}},18^{(8,27)})},29^{(17^{(7,28)}},21^{(9,25)})}), \\ 13^{(24^{(15^{(6,25)}},19^{(8,28)})},26^{(16^{(7,27)}},20^{(9,22)})}). \end{Bmatrix},$$

$$6\begin{Bmatrix} 2^{(0^{(1^{(4,5)}},3^{(8,9)})},7^{(16^{(26,27)}},17^{(28,29)})}), \\ 14^{(22^{(10^{(4,28)}},20^{(9,26)})},23^{(12^{(5,29)}},18^{(8,27)})}), \\ 15^{(24^{(13^{(5,26)}},19^{(8,28)})},25^{(11^{(4,27)}},21^{(9,29)})}). \end{Bmatrix},$$

$$7\begin{Bmatrix} 2^{(0^{(1^{(4,5)}},3^{(8,9)})},6^{(14^{(22,23)}},15^{(24,25)})}), \\ 16^{(26^{(13^{(5,24)}},20^{(9,22)})},27^{(11^{(4,25)}},18^{(8,23)})}), \\ 17^{(28^{(10^{(4,22)}},19^{(8,24)})},29^{(12^{(5,23)}},21^{(9,25)})}). \end{Bmatrix},$$

$$8\begin{Bmatrix} 3^{(0^{(1^{(4,5)}},2^{(6,7)})},9^{(20^{(22,26)}},21^{(25,29)})}), \\ 18^{(23^{(12^{(5,29)}},14^{(6,22)})},27^{(11^{(4,25)}},16^{(7,26)})}), \\ 19^{(24^{(13^{(5,26)}},15^{(6,25)})},28^{(10^{(4,22)}},17^{(7,29)})}). \end{Bmatrix},$$

$$9\begin{Bmatrix} 3^{(0^{(1^{(4,5)}},2^{(6,7)})},8^{(18^{(23,27)}},19^{(24,28)})}), \\ 20^{(22^{(10^{(4,28)}},14^{(6,23)})},26^{(13^{(5,24)}},16^{(7,27)})}), \\ 21^{(25^{(11^{(4,27)}},15^{(6,24)})},29^{(12^{(5,23)}},17^{(7,28)})}). \end{Bmatrix},$$

$$10\begin{Bmatrix} 4^{(1^{(0^{(2,3)}},5^{(12,13)})},11^{(25^{(15,21)}},27^{(16,18)})}), \\ 22^{(14^{(6^{(2,15)}},23^{(12,18)})},20^{(9^{(3,21)}},26^{(13,16)})}), \\ 28^{(17^{(7^{(2,16)}},29^{(12,21)})},19^{(8^{(3,18)}},24^{(13,15)})}). \end{Bmatrix},$$

$$11\begin{Bmatrix} 4^{(1^{(0^{(2,3)}},5^{(12,13)})},10^{(22^{(14,20)}},28^{(17,19)})}), \\ 25^{(15^{(6^{(2,23)}},24^{(24,25)})},21^{(9^{(3,20)}},29^{(12,17)})}), \\ 27^{(16^{(7^{(2,17)}},26^{(13,20)})},18^{(8^{(3,19)}},23^{(12,14)})}). \end{Bmatrix},$$

$$12\begin{Bmatrix} 5^{(1^{(0^{(2,3)}},4^{(10,11)})},13^{(24^{(15,19)}},26^{(16,20)})}), \\ 23^{(14^{(6^{(2,22)}},22^{(24,25)})},18^{(8^{(3,19)}},27^{(11,16)})}), \\ 29^{(17^{(7^{(2,16)}},28^{(10,19)})},21^{(9^{(3,20)}},25^{(11,15)})}). \end{Bmatrix},$$

$$13\begin{Bmatrix} 5^{(1^{(0^{(2,3)}},4^{(10,11)})},12^{(23^{(14,18)}},29^{(17,21)})}), \\ 24^{(15^{(6^{(2,23)}},25^{(24,25)})},19^{(8^{(3,18)}},28^{(10,17)})}), \\ 26^{(16^{(7^{(2,17)}},27^{(11,18)})},20^{(9^{(3,21)}},22^{(10,14)})}). \end{Bmatrix},$$

$$14\begin{pmatrix}6(2^{(0^{(1,3)},7^{(16,17)}},15^{(24^{(13,19)},25^{(11,21)})}),\\22^{(10^{(4,11)},28^{(17,19)}},20^{(9^{(3,21)},26^{(13,16)})}),\\23^{(12^{(5,1,13)},29^{(17,21)}},18^{(8^{(3,19)},27^{(11,16)})}).\end{pmatrix},$$

$$15\begin{pmatrix}6(2^{(0^{(1,3)},7^{(16,17)}},14^{(22^{(10,20)},23^{(12,18)})}),\\24^{(13^{(5,1,12)},26^{(16,20)}},19^{(8^{(3,18)},28^{(10,17)})}),\\25^{(11^{(4,1,10)},27^{(16,18)}},21^{(9^{(3,20)},29^{(12,17)})}).\end{pmatrix},$$

$$16\begin{pmatrix}7(2^{(0^{(1,3)},6^{(14,15)}},17^{(28^{(10,19)},29^{(12,21)})}),\\26^{(13^{(5,1,12)},24^{(15,19)}},20^{(9^{(3,21)},22^{(10,14)})}),\\27^{(11^{(4,1,10)},25^{(15,21)}},18^{(8^{(3,19)},23^{(12,14)})}).\end{pmatrix},$$

$$17\begin{pmatrix}7(2^{(0^{(1,3)},6^{(14,15)}},16^{(26^{(13,20)},27^{(11,18)})}),\\28^{(10^{(4,1,11)},22^{(14,20)}},19^{(8^{(3,18)},24^{(13,15)})}),\\29^{(12^{(5,1,13)},23^{(14,18)}},21^{(9^{(3,20)},25^{(11,15)})}).\end{pmatrix},$$

$$18\begin{pmatrix}8(3^{(0^{(1,2)},9^{(20,21)}},19^{(24^{(13,15)},28^{(10,17)})}),\\23^{(12^{(5,1,13)},29^{(17,21)}},14^{(6^{(2,15)},22^{(10,20)})}),\\27^{(11^{(4,1,10)},25^{(15,21)}},16^{(7^{(2,17)},26^{(13,20)})}).\end{pmatrix},$$

$$19\begin{pmatrix}8(3^{(0^{(1,2)},9^{(20,21)}},18^{(23^{(12,14)},27^{(11,16)})}),\\24^{(13^{(5,1,12)},26^{(16,20)}},15^{(6^{(2,14)},25^{(11,21)})}),\\28^{(10^{(4,1,11)},22^{(14,20)}},17^{(7^{(2,16)},29^{(12,21)})}).\end{pmatrix},$$

$$20\begin{pmatrix}9(3^{(0^{(1,2)},8^{(18,19)}},21^{(25^{(11,15)},29^{(12,17)})}),\\22^{(10^{(4,1,11)},28^{(17,19)}},14^{(6^{(2,15)},23^{(12,18)})}),\\26^{(13^{(5,1,12)},24^{(15,19)}},16^{(7^{(2,17)},27^{(11,18)})}).\end{pmatrix},$$

$$21\begin{pmatrix}9(3^{(0^{(1,2)},8^{(18,19)}},20^{(22^{(10,14)},26^{(13,16)})}),\\25^{(11^{(4,1,10)},27^{(16,18)}},15^{(6^{(2,14)},24^{(13,19)})}),\\29^{(12^{(5,1,13)},23^{(14,18)}},17^{(7^{(2,16)},28^{(10,19)})}).\end{pmatrix},$$

$$22\begin{pmatrix}14(6^{(2^{(0,7)},15^{(24,25)}},23^{(12^{(5,29)},18^{(8,27)})}),\\10^{(4^{(1,0,5)},11^{(25,27)}},28^{(17^{(7,29)},19^{(8,24)})}),\\20^{(9^{(3^{(0,8)},21^{(25,29)}},26^{(13^{(5,24)},16^{(7,27)})}).\end{pmatrix},$$

$$23\begin{pmatrix}14(6^{(2^{(0,7)},15^{(24,25)}},22^{(10^{(4,28)},20^{(9,26)})}),\\12^{(5^{(1^{(0,4)},13^{(24,26)}},29^{(17^{(7,28)},21^{(9,25)})}),\\18^{(8^{(3^{(0,9)},19^{(24,28)}},27^{(11^{(4,25)},16^{(7,26)})}).\end{pmatrix},$$

$$24\begin{pmatrix}13(5^{(1^{(0,4)},12^{(23,29)}},26^{(16^{(7,27)},20^{(9,22)})}),\\15^{(6^{(2^{(0,7)},14^{(22,23)}},25^{(11^{(4,27)},21^{(9,29)})}),\\19^{(8^{(3^{(0,9)},18^{(23,27)}},28^{(10^{(4,22)},17^{(7,29)})}).\end{pmatrix},$$

$$25\begin{pmatrix}11(4^{(1^{(0,5)},10^{(22,28)}},27^{(16^{(7,26)},18^{(8,23)})}),\\15^{(6^{(2^{(0,7)},14^{(22,23)}},24^{(13^{(5,26)},19^{(8,28)})}),\\21^{(9^{(3^{(0,8)},20^{(22,26)}},29^{(12^{(5,23)},17^{(7,28)})}).\end{pmatrix},$$

$$26\begin{pmatrix}13(5^{(1^{(0,4)},12^{(23,29)}},24^{(15^{(6,25)},19^{(8,28)})}),\\16^{(7^{(2^{(0,6)},17^{(28,29)}},27^{(11^{(4,25)},18^{(8,23)})}),\\20^{(9^{(3^{(0,8)},21^{(25,29)}},22^{(10^{(4,28)},14^{(6,23)})}).\end{pmatrix},$$

$$27\begin{pmatrix}11(4^{(1^{(0,5)},10^{(22,28)}},25^{(15^{(6,24)},21^{(9,29)})}),\\16^{(7^{(2^{(0,6)},17^{(28,29)}},26^{(13^{(5,24)},20^{(9,22)})}),\\18^{(8^{(3^{(0,9)},19^{(24,28)}},23^{(12^{(5,29)},14^{(6,22)})}).\end{pmatrix},$$

$$28\begin{pmatrix}10(4^{(1^{(0,5)},11^{(25,27)}},22^{(14^{(6,23)},20^{(9,26)})}),\\17^{(7^{(2^{(0,6)},16^{(26,27)}},29^{(12^{(5,23)},21^{(9,25)})}),\\19^{(8^{(3^{(0,9)},18^{(23,27)}},24^{(13^{(5,26)},15^{(6,25)})}).\end{pmatrix},$$

$$29\begin{pmatrix}12(5^{(1^{(0,4)},13^{(24,26)}},23^{(14^{(6,22)},18^{(8,27)})}),\\17^{(7^{(2^{(0,6)},16^{(26,27)}},28^{(10^{(4,22)},19^{(8,24)})}),\\21^{(9^{(3^{(0,8)},20^{(22,26)}},25^{(11^{(4,27)},15^{(6,24)})}).\end{pmatrix},$$

# ПРИЛОЖЕНИЕ П2

Ниже приведены проекции 15(4,2)-компактного графа с образующим 3-цикл вершинами {0, 1, 2}. Выделение вершин в проекциях аналогично описанному в приложении П1.

$C_0 = 3003$, $C_{0.1} = 170$

Решение системы проекций здесь достигнуто за 9 шагов:

1-й шаг: добавлено ребро 3-14

2-й шаг: добавлено ребро 4-11

3-й шаг: добавлено ребро 3-12

4-й шаг: добавлено ребро 4-13

5-й шаг: добавлено ребро 5-14

6-й шаг: добавлено ребро 6-9. Получены 3 ребра: 5-10, 6-13, 13-14

7-й шаг: добавлены ребра 5-10, 6-13, 13-14, полученные на предыдущем шаге

8-й шаг: добавлено ребро 7-10, полученное на шаге 7. Получено ребро 8-11

9-й шаг: добавлено ребро 8-11, полученное на шаге 8. На этом шаге получены все 3 оставшиеся неизвестными ребра: 8-12, 9-11, 10-13

$$0\begin{Bmatrix}1^{(2,5,6)},\\2^{(1,7,8)},\\3^{(9,12,14)},\\4^{(10,11,13)}.\end{Bmatrix}, \quad 1\begin{Bmatrix}0^{(2,3,4)},\\2^{(0,7,8)},\\5^{(10,11,14)},\\6^{(9,12,13)}.\end{Bmatrix}, \quad 2\begin{Bmatrix}0^{(1,3,4)},\\1^{(0,5,6)},\\7^{(10,9,13)},\\8^{(11,12,14)}.\end{Bmatrix},$$

$$3\begin{Bmatrix}0^{(1,2,4)},\\9^{(6,7,11)},\\12^{(6,8,10)},\\14^{(5,8,13)}.\end{Bmatrix}, \quad 4\begin{Bmatrix}0^{(1,2,3)},\\10^{(5,7,12)},\\11^{(5,8,9)},\\13^{(6,7,14)}.\end{Bmatrix}, \quad 5\begin{Bmatrix}1^{(0,2,6)},\\10^{(4,7,12)},\\11^{(4,8,9)},\\14^{(3,8,13)}.\end{Bmatrix},$$

$$6\begin{Bmatrix}1^{(0,2,5)},\\9^{(3,7,11)},\\12^{(3,8,10)},\\13^{(4,7,14)}.\end{Bmatrix}, \quad 7\begin{Bmatrix}2^{(0,1,8)},\\9^{(3,6,11)},\\10^{(4,5,12)},\\13^{(4,6,14)}.\end{Bmatrix}, \quad 8\begin{Bmatrix}2^{(0,1,7)},\\11^{(4,5,9)},\\12^{(3,6,10)},\\14^{(3,5,13)}.\end{Bmatrix},$$

$$9\begin{Bmatrix}3^{(0,12,14)},\\6^{(1,12,13)},\\7^{(2,10,13)},\\11^{(4,5,8)}.\end{Bmatrix}, \quad 10\begin{Bmatrix}4^{(0,11,13)},\\5^{(1,11,14)},\\7^{(2,9,13)},\\12^{(3,6,8)}.\end{Bmatrix}, \quad 11\begin{Bmatrix}4^{(0,10,13)},\\5^{(1,10,14)},\\8^{(2,12,14)},\\9^{(3,6,7)}.\end{Bmatrix},$$

$$12\begin{Bmatrix}3^{(0,9,14)},\\6^{(1,9,13)},\\8^{(2,11,14)},\\10^{(4,5,7)}.\end{Bmatrix}, \quad 13\begin{Bmatrix}4^{(0,10,11)},\\6^{(1,9,12)},\\7^{(2,10,9)},\\14^{(3,5,8)}.\end{Bmatrix}, \quad 14\begin{Bmatrix}3^{(0,9,12)},\\5^{(1,10,11)},\\8^{(2,11,12)},\\13^{(4,6,7)}.\end{Bmatrix},$$